\documentclass{emulateapj}
\usepackage{verbatim}
\usepackage{natbib}
\usepackage{amsmath}
\usepackage{ulsy}
\usepackage{subfigure}
\newcommand{\pfrac}[2]{\left(\frac{#1}{#2}\right)}

\newcommand{\vrel}{v_{\rm rel}}
\newcommand{\Rcap}{R_{\rm cap}}
\newcommand{\LIPAD}{{\tt LIPAD}}

\shorttitle{Challenges in Forming Giant Planet Cores via Pebble Accretion}
\shortauthors{Kretke \& Levison}

\begin{document}

\title{Challenges in Forming the Solar System's Giant Planet Cores via Pebble Accretion}
\author{K. A. Kretke$^{1}$ and H. F. Levison$^{1}$}
\affil{$^1$Southwest Research Institute, 1050 Walnut St, Suite 300, Boulder, CO 80302, USA}
\email{kretke@boulder.swri.edu}
\keywords{Planets and Satellites: Formation}

\begin{abstract}
	Though $\sim10 M_\oplus$ mass rocky/icy cores are commonly held as a prerequisite for the formation of gas giants, theoretical models still struggle to explain how these embryos can form within the lifetimes of gaseous circumstellar disks. In recent years, aerodynamic-aided accretion of ``pebbles,'' objects ranging from centimeters to meters in size, has been suggested as a potential solution to this long-standing problem. 
While pebble accretion has been demonstrated to be extremely effective in local simulations that look at the detailed behavior of these pebbles in the vicinity of a single planetary embryo, to date there have been no global simulations demonstrating the effectiveness of pebble accretion in a more  complicated, multi-planet environment.
Therefore, we have incorporated the aerodynamic-aided accretion physics into \LIPAD, a Lagrangian code that can follow the collisional / accretional / dynamical evolution of a protoplanetary system, to investigate how pebble accretion manifests itself in the larger planet formation picture. 
We find that under generic circumstances, pebble accretion naturally leads to an ``oligarchic'' type of growth in which a large number of planetesimals grow to similar-sized planets.  In particular, our simulations tend to form hundreds of Mars- and Earth-mass objects between 4 and 10 AU.  While the merging of some oligarchs may allow them to grow massive enough to form giant planet cores, 
leftover oligarchs lead to planetary systems that cannot be consistent with our own solar system.  We investigate various ideas presented in the literature (including evaporation fronts and planet traps) and find that none easily overcome this tendency toward oligarchic growth.

\end{abstract}

\section{Introduction}\label{sec:intro}
The core accretion model \citep{Pollack.etal.1996} has been the most successful model for giant planet formation for almost 20 yr now, and yet it still faces significant challenges.
According to this model, giant planets emerge in two steps: first a massive rocky/icy core forms and only then can a hydrogen/helium envelope be accreted from the surrounding gaseous protoplanetary disk.
This solid core must form quickly, within the few-million-year lifetime of the gas disk \citep{Haisch.etal.2001,Hernandez.etal.2008,Williams.2012}.
The exact required mass of this core is still under active debate, but values around 10 Earth masses ($M_\oplus$) are generally invoked (although it is possible that masses as low as 5 $M_\oplus$ may be sufficient; \citet{Hori.Ikoma.2011}).

Unfortunately, it is quite challenging to build 10 $M_\oplus$ cores quickly using standard models \citep{Ngo.etal.2012}.
The crux of the issue is that, as a proto-core grows, it excites the eccentricities and inclinations of surrounding planetesimals.  
This dynamical excitement increases the relative velocity of close encounters, decreasing the effectiveness of gravitational focusing and thus the likelihood of actual collisions.
The result is core formation timescales that are too long.
In order to get around this problem, some authors have suggested that, if planetesimals are small, then the eccentricities will be effectively damped as a result of either mutual collisions or gas drag \citep[e.g.][]{Goldreich.etal.2004a}.
This damping yields dynamically cold planetesimals that are strongly affected by gravitational focusing and hence should have large capture cross sections \citep[e.g.][; Figure \ref{fig:cross-section}]{Rafikov.2004}.
Unfortunately, if planetesimals lose sufficient energy to stay dynamically cold, then this is the same condition for the planets to open gaps in the planetesimal disk \citep{Levison.etal.2010a}.  Accretion will stop.  
Therefore we are left with a conundrum; neither large nor small planetesimals seem to be able to effectively form giant planet cores.

However, if one looks at even smaller objects, pebbles or boulders with stopping times ($t_s$) comparable to the orbital period, a possible solution emerges.  
These objects (which we call pebbles as a shorthand) are very effectively damped by gas drag so that they will be dynamically quite cold.  This drag also causes these pebbles to drift extremely rapidly in a normal pressure-supported gaseous disk \citep{Adachi.etal.1976,Weidenschilling.1977}.  As the pebble's drift timescale is much shorter than nearby a planet's synodic period, the planet cannot effectively open gaps in the pebble disk.
Now, if an object is too small (its stopping time is short compared to the time for it to encounter the planetesimal) then it will be so well coupled to the gas that it follows the gas and flows around the planetesimal and will not be efficiently accreted.
However, if the stopping time is comparable to the encounter time then the particle can be deflected out of the gas stream line and thereby lose enough energy during the encounter to be gravitationally captured by the planetesimal.
This means that as these pebbles drift by a planetesimal they will spiral inward and can be accreted very efficiently, with accretion cross sections as large as the entire Hill sphere \citep[][hereafter OK10]{Johansen.Lacerda.2010,Ormel.Klahr.2010}.

Furthermore, there may be good reason to believe that the early protoplanetary disk was in fact filled with these types of pebbles.  While binary collisions appear effective to grow dust grains from micron to pebble sizes, once the solids reach pebble size their growth is stifled for three reasons.  First, as particles grow larger their sticking efficiency is reduced, possibly leading to a regime in which collisions are more likely to lead to ``bouncing'' rather than accretion \citep{Zsom.etal.2010}.  Additionally, as particles grow larger they can attain larger relative velocities, increasing the likelihood that collisions will be destructive \citep{Blum.Wurm.2000,Blum.Wurm.2008}.  Furthermore, if the do grow then, as discussed above, they will drift inward very rapidly.  

However, given enough of these pebble-sized objects, one can expect interactions between the solids and the gas \citep[such as the streaming instability][]{Youdin.Goodman.2005,Youdin.Johansen.2007} to concentrate pebbles into clumps that can become gravitationally unstable, directly forming large planetesimals \citep{Johansen.etal.2007}.
If one looks at the present-day asteroid belt and Kuiper Belt there is evidence that the planetesimals were in fact originally somewhere around 100 km in size. \citep{Morbidelli.etal.2009,Nesvorny.etal.2010}.  Therefore, it seems plausible that the early solar system consisted only of large ($>100$ km-sized) planetesimals and a sea of small pebbles. 

Indeed, if a substantial fraction of the disk's solids are in pebbles, then as the pebbles rapidly drift past planetesimals with very large capture cross sections, the accretion rate can be extremely high.
\citet[][hereafter LJ12]{Lambrechts.Johansen.2012} coined the term ``pebble accretion'' to describe this interesting process in which pebbles are swept up by existing planetesimals and postulated that this very rapid accretion could lead to the formation of the cores of giant planets.

In this first of a series of papers we investigate the feasibility of forming the giant planets in our solar system from an initial population of planetesimals and pebbles.
This work builds on the seminal works of OK10 and LJ12 which presented, in detail, how pebbles are accreted by a single planetesimal or protoplanet.
These researchers have found that pebble accretion can be very efficient, making it a promising mechanism to form the cores of giant planets.  
Indeed, LJ2012 postulated that if there were only a couple of planetesimals large enough to efficiently accrete pebbles, then those planetesimals could directly form the giant planet cores. 
However, these models have not yet demonstrated whether pebble accretion can be the dominant mechanism of planet growth in a more comprehensive planet formation calculation with many interacting planetesimals.

In this first paper we are interested in investigating a relatively simple
scenario, inspired by the aforementioned theoretical work and designed to focus
on the process of planetesimal formation by the streaming instability followed
by pebble accretion.  In this scenario we assume that dust grows effectively
and efficiently up to pebble size, leading to a large pebble-to-gas ratio in
which a substantial fraction of the solids in the disk are in the form of
pebbles.  These pebbles then settle toward the midplane and, owing to the
oww
streaming instability, form clumps that become gravitationally unstable
\citep{Youdin.Goodman.2005, Johansen.etal.2007}.  Then, because this
planetesimal formation is not efficient, there will be a large
population of remnant pebbles \citep{Johansen.etal.2009}.  These pebbles will
drift inward and be accreted by the planetesimals, leading to the formation of
planets (LJ2012).  We furthermore assume that the large pebble-to-gas ratio
required by the streaming instability is difficult to achieve and therefore
only occurs once in the disk lifetime. As such, we will model a single epoch of
pebble formation followed rapidly by a single epoch of planetesimal formation.
After the remnant pebbles have either drifted out of the disk or been accreted,
we assume that not enough mass is ever again in pebbles in order to impact the
planet formation process, through either planetesimal formation or pebble
accretion.  While this model is simplistic and perhaps na\:ive, as it does not
address the challenge of how this large mass of pebbles would have formed in a
timescale short compared to the drift timescale, it yields initial conditions
that will allow us to directly investigate the problem of pebble accretion as
a natural outcome of the leftovers of planetesimal formation.

We note that this work differs from recent work by \citet{Chambers.2014} in that we are interested in investigating the initial proposal by LJ12 that pebble accretion can lead directly from planetesimals to giant planet cores.  \citeauthor{Chambers.2014} found that, under certain conditions, pebble accretion may enhance the subsequent planetary growth and evolution if one already has reached the ``classical'' oligarchic growth stage in planet formation via accretion of planets from planetesimals \citet{Kokubo.Ida.1998,Kokubo.Ida.2000}, but did not address the evolution of a system in which pebbles are important from the very beginning.

In this paper we utilize the results of OK10 and LJ12 to investigate how planetary systems would evolve from an initial population of pebbles and planetesimals through the formation of giant planets.
We begin in Section \ref{sec:LIPAD} by discussing the numerical planet formation code we use in this paper.
We then present a short review of the physical processes behind pebble accretion in Section \ref{sec:review}.
We demonstrate the implementation of pebble accretion into our code in Section \ref{sec:initial} and then 
present our fiducial model in Section \ref{sec:fiducial}.
In Section \ref{sec:analytics} we briefly discuss the analytical reasons behind the behavior seen in this calculation.
In Section \ref{sec:eta_size} we validate the analytical discussion by demonstrating how the disk structure and pebble size impact our numerical results.
In Section \ref{sec:planetesimals} we discuss the importance of the initial planetesimal population.
In Section \ref{sec:other} we then turn to other ideas that have been suggested in the literature to aid core formation with pebbles: shielding,  evaporation fronts, and planet traps.
In Section \ref{sec:discussion} we discuss the caveats and implications of our results, and summarize our conclusions.

\section{Description of our Numerical Model}\label{sec:LIPAD}
For this work we have employed a modified version of \LIPAD\ ({\bf L}agrangian {\bf I}ntegrator for {\bf P}lanetary
{\bf A}ccretion and {\bf D}ynamics),
a particle-based Lagrangian code that
can follow the dynamical/collisional/accretional evolution of a large
number of planetesimals through the entire growth process to
become planets.
For full details of the code and extensive test suites see \citet{Levison.etal.2012}, but we summarize the most relevant attributes of the code here.

\LIPAD\ is built on top of the N-body integrator SyMBA
\citep{Duncan.etal.1998}.  In order to handle the very large number of
sub-kilometer objects required by many simulations, \LIPAD\ utilizes tracer
particles. Each tracer represents a large number of small bodies with roughly
the same orbit and size, and is characterized by three numbers: the physical
radius, the bulk density, and the total mass of the disk particles represented
by the tracer.  \LIPAD\ employs statistical algorithms that follow the
dynamical and collisional interactions between the tracers. When a tracer is
determined to have been struck by another tracer, it is assigned a new radius
according to the probabilistic outcome of the collision based on a
fragmentation law \citep[][using the ice parameters $Q_0 = 7\times 10^7\,{\rm
erg\,g}^{-1}$, $B=2.1\,\rm{erg\,cm}^3\,\rm{g}^{-2}$, $a=-0.45$ and
$b=1.19$]{Benz.Asphaug.1999}.  This way, the conglomeration of tracers
represents the size distribution of the evolving planetesimal population.  In
this work, we do not allow our pebbles to either grow or fragment; therefore,
particles below 1~km in size are not involved in the collisional cascade.
\LIPAD\ also includes statistical algorithms for viscous stirring, dynamical
friction, and collisional damping among the tracers. The tracers mainly
dynamically interact with the larger planetary-mass objects via the normal
$N$-body routines, which naturally follow changes in the trajectory of tracers
due to the gravitational effects of the planets and {\it vice versa}.  \LIPAD\
is therefore unique in its ability to accurately handle the mixing and
redistribution of material due to gravitational encounters, including gap
opening, and resonant trapping, while also following the fragmentation and
growth of bodies.  Thus, it is well suited to follow the collisional cascade of
a population of planetesimals while they gravitationally interact with a system
of planets.  

For this work we have made a number of optional additions to \LIPAD.
While in principle the basic \LIPAD\ engine can treat the interaction and accretion of pebbles, calculating the trajectories of incoming pebbles as they interact with the gas around the (proto-)planets is not computationally practical in these large-scale calculations.
In order to significantly increase the computational speed of the calculation, 
instead of directly integrating the trajectory of a pebble as it spirals toward the planet, if a pebble is within a planet's Hill sphere ($R_H$) and satisfies the criteria found by OK10 (which will be described in Equation~\ref{eq:Rcap}) then we allow the pebble to be accreted.
Additionally, although the pebbles can be accreted by
planetesimals, we ignore collisions between pebbles.

Another optional addition to \LIPAD\ is gas turbulence.  
In the absence of turbulence, aerodynamic drag causes these small pebbles to rapidly concentrate at the midplane.  
Not only are there many mechanisms to drive turbulence in disks \citep[e.g.][]{Balbus.2000,Lithwick.2009,Vorobyov.Basu.2008}, but even in the absence of external forcing, as the pebbles settle, they can self-stir the gas
disk either by vertical shearing \citep{Goldreich.Ward.1973, Cuzzi.etal.1993} or by streaming instabilities \citep{Youdin.Goodman.2005, Youdin.Johansen.2007, Johansen.Youdin.2007}.
We are not modeling the detailed behavior of the gas; therefore, we have added random kicks to the pebbles in order to simulate the effects of turbulence.  

Given our resolution, we are unable to resolve any correlated behavior of pebbles due to turbulence; therefore, we use a simple prescription inspired by \citet{Youdin.Lithwick.2007} and give the pebbles random kicks of strength
\begin{equation}
	\delta v = h_p \Omega_K \sqrt{\frac{\delta t}{t_s}}. \nonumber
\end{equation}
Here $h_p$ is the desired particle scale height and $\Omega_K$ is the Keplerian orbital frequency.  
This is equivalent to assuming that the correlation time for eddies is the same as our time step ($\delta t$).
For our fiducial model we choose $h_p$ such that the pebble density in the midplane is equal to the gas density (consistent with the streaming instability).  
Following \citet{Youdin.Lithwick.2007}, we can relate this $\delta v$ to the common \citet{Shakura.Sunyaev.1973} $\alpha$ parameter.  For particles near $\tau=1$, if the solid-to-gas ratio is 0.01, this corresponds to an $\alpha \approx 10^{-4}$.
As described in \citet{Levison.etal.2012}, we calculate the aerodynamic drag on the particles using the formalism of \citet{Adachi.etal.1976}.

As in previous LIPAD runs, we have used the prescription in \citet{Inaba.Ikoma.2003} to include the enhanced capture cross section due to aerodynamic drag. 
Additionally, as we are interested in the gross evolution of a system after the formation of a potential giant planet core.  Therefore, we have added in a simple optional prescription  allowing cores to accrete gas envelopes. 
In order to accrete gas, the core size must be above a critical value, which depends on the mass accretion rate of solids onto the core.
We follow \citet{Rafikov.2006} to determine whether the current core mass is above the critical value given the current mass accretion rate.
If this criterion is met, then we grow the planet of mass $M_p$ by allowing gas to accrete on the Kelvin-Helmholtz timescale ($t_{\rm KH}$), so that
\begin{equation}
	\dot M_g = \frac{M_p}{t_{\rm KH}}. \nonumber
\end{equation}
We follow \citep{Ida.Lin.2008a} and approximate this timescale as 
\begin{equation}
	t_{\rm KH} = 10^9 \pfrac{M_p}{M_\oplus}^{-3} {\rm yr}. \nonumber
\end{equation}
We limit gas accretion to the Bondi accretion rate,
\begin{equation}
	\dot M_{\rm g, max} = \frac{4\pi\rho_g G^2 M_p^2}{c_s^3}, \nonumber
\end{equation}
where $\rho_g$ is the gas density and $c_s$ is the local sound speed.
We arbitrarily cut off gas accretion when a planet reaches one Jupiter mass.

Additionally, once a giant planet begins to grow we allow it to open up a gap in the gas disk.  A planet opens a gap if
\begin{equation}
	\frac{3}{4} \frac{h}{R_H} + \frac{5 M_\odot}{M_p\mathcal{R}} < 1, \nonumber
\end{equation}
where $\mathcal{R} = (v_K/c_s)^2/\alpha$ \citep{Crida.etal.2006}, $h$ is the gas scale height, $M_\odot$ is the mass of the Sun, and $v_K$ is the Keplerian velocity.
We follow \citet{Varniere.etal.2004} and assume that the half-width of this gap is
\begin{equation}
	\Delta = 0.29 \pfrac{M_p}{M_*}^{2/3} \mathcal{R}^{1/3} r_p. \nonumber
\end{equation}
where $r_p$ is the distance from the Sun to the planet.  Inside this region we turn off aerodynamic drag, type I migration, and eccentricity damping.

In most of the simulations presented here we do not allow the growing planets to migrate via type I migration (with the exception of the planet trap simulations in Section \ref{sec:trap}), although we do include type-I eccentricity damping \citep{Papaloizou.Larwood.2000}.  In Section \ref{sec:trap} we present the details for our type I migration for those runs.  In all runs we neglect type II migration.

\section{Pebble Accretion Rates}\label{sec:review}
The physics behind pebble accretion has been discussed in detail in OK10 and LJ12.  We encourage the interested reader to find full details within those two papers, but, for convenience, we summarize the key points here.

In order for gas drag to significantly enhance the probability that a particle is accreted by a target object, two criteria must be met.  First, the particle's stopping timescale must be long compared to the timescale for the particle to be deflected by the target object's gravity, or else it is too well coupled to the gas to be accreted. 
Second, the particle's stopping time must be short compared to the time for it to drift past the target, or else the gas drag is not important in the accretion process.

Relevant to the first criteria, OK10 found that, specifically, the gravitational encounter timescale must be shorter than four times the stopping time, i.e.
\begin{equation}
	\vrel\frac{b^2}{G M_p} < 4 t_s, \nonumber
\end{equation}
where $b$ is the impact parameter, $G$ is the gravitational constant, $M_p$ is the mass of the growing planetesimal, and $v_{\rm rel}$ is the speed at which the particle approaches the planetesimal.
This means that in order for a pebble to be captured, it must approach at a distance of 
\begin{equation}
	b < \tilde R_C \equiv \pfrac{4 G M_p t_s}{\vrel}^{1/2}.
	\label{eq:R_C}
\end{equation}

Relevant to the second criterion,
in the two-body problem a particle is deflected by 90 deg if its impact parameter is 
\begin{equation}
b = b_{90} \equiv \frac{G M_p}{\vrel^2} \nonumber
\end{equation}
(this is called the Bondi radius in LJ2012, but we distinguish it from the canonical Bondi radius for gas accretion).  Let us define the critical crossing timescale as the time for a particle to cross this 90 deg scattering radius
\begin{equation}
	t_{s,*} \equiv \frac{b_{90}}{\vrel} = \frac{G M_p}{\vrel^3}. \nonumber
\end{equation}
This is effectively the time that a particle spends strongly interacting with the target.  If $t_s \gg t_{s,*}$ then the particle does not feel the gas drag, and the actual capture cross section is much smaller than the limit given by Equation (\ref{eq:R_C}).
OK10 found that a pebble is efficiently accreted only if 
\begin{equation}
	b < R_C \equiv \tilde R_C \exp\left[-\pfrac{t_s}{4 t_{s,*}}^{\gamma}\right].
\label{eq:Rcap}
\end{equation}
where $\gamma = 0.65$.  This provides quite a sharp cutoff in the mass accretion rate, so that small targets grow only as a result of direct hits, while larger ones readily grow.

In our particle-based numerical model (described in Section \ref{sec:LIPAD}) we employ Equation (\ref{eq:Rcap}) to determine when a pebble is accreted by a growing planetesimal or planet. However, we can also integrate over $R_C$ and calculate the expected pebble accretion rate
\begin{equation}
	\dot{M}_p = \int_{s_{\rm min}}^{s_{\rm max}} \int_0^{2\pi} \int_0^{R_C(s,M_p)}  \frac{d \rho(s,z)}{ds} v_{\rm rel}(s,x) b~db~d\theta~ds,
	\label{eq:Mdot}
\end{equation}
where $x$ and $z$ are the local Cartesian coordinates in the radial and vertical directions, respectively, and $\theta\equiv \arctan (z/x)$, so that $z=b\sin\theta$ and $x=b\cos\theta$. $R_C(s,M_p)$ is the capture radius which depends on the size of the particles and the relative velocity of those particles.

Aerodynamic drag between pebbles and the pressure-supported disk gas causes pebbles to drift \citep{Adachi.etal.1976,Weidenschilling.1977}.  
The relative velocity of a pebble drifting by at a radial distance, $x$, from a planetesimal can be approximated as a sum of this aerodynamic driven drift and Keplerian shear, 
\begin{equation}
\vrel \approx \left(\Theta \eta a + x\right)\Omega_K.\label{eq:vrel}
\end{equation}
Here $a$ is the planetesimal's semimajor axis, $\Omega_K\equiv\sqrt{G M_*/a^3}$ is the Keplerian orbital frequency, and the parameters $\Theta$ and $\eta$ parameterize the amount of pebble drift.  We define these in the next two paragraphs.

Due to the pressure ($P$) gradient in the disk, the gas generally rotates at a slightly sub-Keplerian velocity of
\begin{equation}
	v_{\rm cir} = v_K (1-\eta) \nonumber
\end{equation}
where
\begin{equation}
\eta = -\frac{1}{2}\pfrac{h}{r}^2\frac{d\ln P}{d\ln r}. \nonumber
\end{equation}
A power-law disk with a surface density profile of
\begin{equation}
	\Sigma = \Sigma_0 r_{\rm AU}^{p} \nonumber
\end{equation}
(where $r_{\rm AU}\equiv r/(1 \rm{AU})$) and a disk scale height of 
\begin{equation}
	h = h_0 r_{\rm AU}^{q} \nonumber
\end{equation}
has
\begin{equation}
	\eta = \frac{3-p-q}{2}\left(\frac{h_0}{\rm AU} r_{\rm AU}^{q-1}\right)^2. \nonumber
\end{equation}
For reference, in the minimum-mass solar nebula \citep[MMSN,][]{Weidenschilling.1977a,Hayashi.1981}
$\eta = 2\times 10^{-3} r_{\rm AU}^{1/2}$.

The pebbles have a motion different from both the gas and a Keplerian orbit.
It is convenient to parameterize the pebbles using the non-dimensional Stoke's number $\tau \equiv t_s \Omega_K.$ 
The gas drag causes the pebbles to drift relative to a Keplerian orbit, in both the azimuthal and radial directions.
The net relative velocity in Equation~(\ref{eq:vrel}) can be found in the laminar case using
\begin{equation}
\Theta =  \frac{\sqrt{4\tau^2+1}}{\tau^2+1} \nonumber
\end{equation}
(LJ2012).  For pebbles with $\tau \lesssim 1$, $\Theta$ can be well approximated by unity.

If the relative velocity is given by Equation (\ref{eq:vrel}), the capture radius according to the first criteria can be found analytically by solving the cubic equation
\begin{equation}
\tilde R^3 + \frac{2}{3}\Theta\eta a \tilde R^2 - 8 R_H^3 = 0.
\label{eq:R}
\end{equation}
In the shear dominated regime (for larger planets), capture can only occur if
\begin{equation}
	b < \tilde R_1 \equiv 2 R_H \tau^{1/3}, 
	\label{eq:R_shear}
\end{equation}
while in the head wind dominated regime (for smaller planets), capture can only occur if
\begin{equation}
	b < \tilde R_2 \equiv \sqrt\frac{12 R_H^3}{\Theta\eta a}.
	\label{eq:R_headwind}
\end{equation}
However, for small planets (or planetesimals), the second criteria (and hence the exponential part of Equation (\ref{eq:Rcap})) is normally more restrictive.

\begin{figure} 
	\includegraphics[width=\columnwidth]{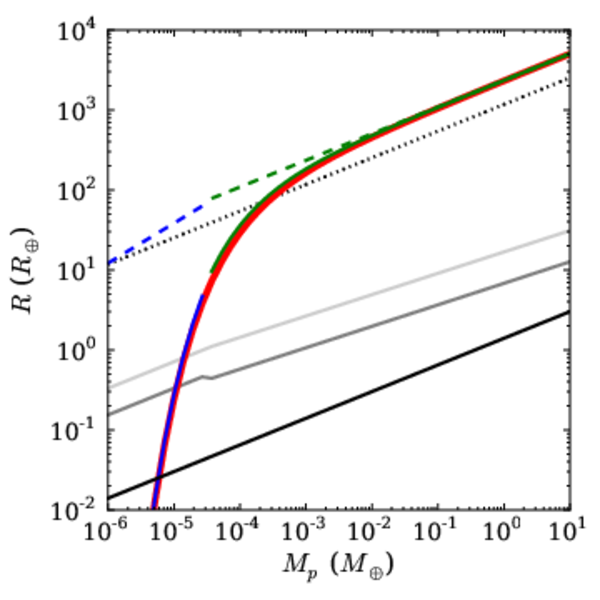}
	\caption{
		Comparisons of the different capture radii discussed in this problem.  The thick red line shows the capture radius for a $\tau=1$ pebble as a function of planet mass at 5 AU in the fiducial disk.  The green and blue dashed curves show Equations~(\ref{eq:R_shear}) and (\ref{eq:R_headwind}), respectively, while the green and blue solid curves show the capture radii when modified by Equation~(\ref{eq:Rcap}).  The black line indicates the physical cross section of the planet, while the dotted black line marks the Hill sphere.  For reference, we add approximations to the capture radius for planetesimals of size 100 km and 1 km as gray curves (dark and light, respectively) as given in \citet{Rafikov.2004}.
	\label{fig:cross-section}}
\end{figure}

For reference, in Figure \ref{fig:cross-section} we show example capture radii as a function of planet mass in our fiducial disk model (described in Section \ref{sec:fiducial}).
Note that for planetesimals larger than $10^{-5} M_\oplus$ the effective capture radius in pebble accretion is substantially larger than the comparable capture radius for planetesimal accretion.

\section{Initial Tests} \label{sec:initial}
\begin{figure} 
\includegraphics[width=\columnwidth]{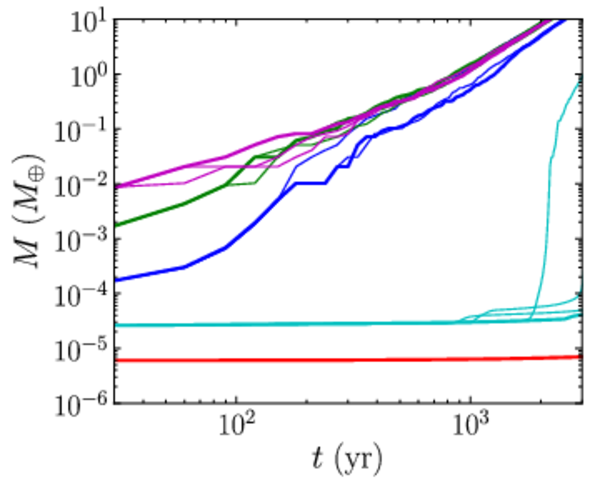}
	\caption{Growth of planetesimals of initial radius 200, 300, 500, 1000, and 2000 km (red, cyan, blue, green and magenta, respectively). We show multiple runs for each initial planetesimal size to demonstrate the stochastic nature of \LIPAD. \label{fig:single}}
\end{figure}

\begin{figure} 
\includegraphics[width=\columnwidth]{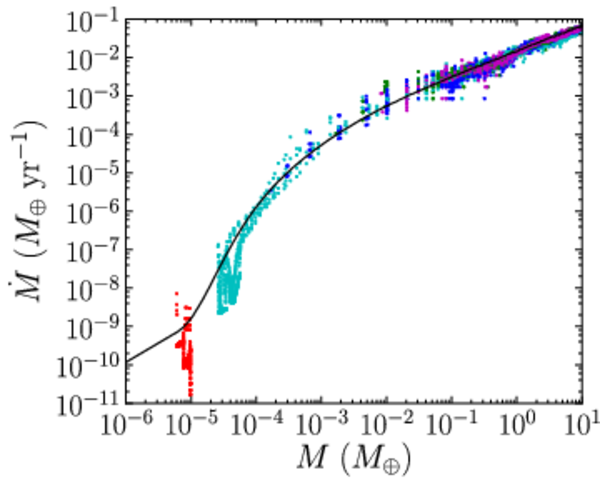}
\caption{Pebble accretion rate as a function of mass from the \LIPAD\ simulations overlaid on the results of Equation (\ref{eq:Mdot}) (with the \citet{Ormel.Klahr.2010} hyperbolic regime cross section for low-mass planetesimals).
	\label{fig:single_Mdot}
	} 
\end{figure}

To demonstrate the efficiency of pebble accretion to produce large objects quickly, we first place a single embryo at 4.5 AU in a sea of pebbles and allow it to accrete pebbles that drift inward from large heliocentric distances.
We assume a surface density for the gas disk of $\Sigma=4500 r_{\rm AU}^{-1}$, which is about 2.5 times the MMSN at 1 AU but falls off more shallowly.  The gas disk has a scale height of $h=0.035 r_{\rm AU}^{5/4} {\rm AU}$.
Pebbles are chosen to have Stoke's numbers near unity at the location of our planetesimal.
We assume solar composition so that the solids (assumed to be all incorporated into pebbles) are 1\% of the mass of the gas \citep[consistent with][]{Lodders.2003}. We set the turbulence so that the initial density of pebbles at the midplane equals the midplane gas density. 

In Figure \ref{fig:single} we show the growth of an individual planetesimal as a function of time.  One can see that if an initial planetesimal is small then there is negligible growth over the time frame shown.  However, for initially large planetesimals there is exceptionally rapid growth; a 500 km planetesimal can grow to 10 $M_\oplus$ in 2000 yr!
In order to roughly maintain the pebble-to-gas ratio in these simulations as pebbles drift inward, we add new pebbles to the outer edge of the computational domain at a rate of $\dot M_{\rm drift} = 0.22 M_{\oplus}~{\rm yr}^{-1}$. The pebbles continue to enter the domain for the length of the simulation.
In total, over 650 $M_\oplus$ of pebbles drifted to the planetesimal during the simulations.  This demonstrates a key aspect of pebble accretion: as long as enough pebbles are present in the disk, a single planetesimal (of sufficient size) can grow extraordinarily rapidly.

We show multiple runs at each initial planetesimal size so that the stochastic nature of pebble accretion in \LIPAD\ is apparent.  For most of the runs the growth rates are similar between planetesimals with the same initial mass.  The most notable exception to this is one of the initial 300 km particles.  In this run the planetesimal happens to accrete a large number of pebbles just before the 2000 yr mark, and is able to grow to 1 $M_\oplus$. This has to do with the very steep change in the accretion radius for planetesimals in this size range (see Figure \ref{fig:cross-section}).  The planetesimal happened to grow large enough that its growth rate became significantly larger than other, slightly smaller planetesimals.  While this stochastic nature arises from our use of tracer particles, we believe that it is likely representative of real, undoubtedly somewhat inhomogeneous, protoplanetary disks. 

In Figure \ref{fig:single_Mdot} we show the mass accretion rate onto the planetesimals as a function of mass for the above simulations.  We compare this to what would be expected by the analytical model given in Equation (\ref{eq:Mdot}) if the relative velocities are assumed to be those of a laminar disk (Equation (\ref{eq:vrel})).  The stochastic nature is apparent, but these simulations demonstrate that at a given planetesimal mass (and for our chosen level of turbulence) the mass accretion rate is well determined, and is dominated by the laminar, not the turbulent, components of the relative velocities.  This means that while turbulence is very important for determining the particle scale height, it does not dominate the encounter velocity.  Additionally, it is clear that \LIPAD\ is able to successfully reproduce the behavior described in OK10.

\section{Fiducial Model}\label{sec:fiducial}

\begin{figure} 
	\includegraphics[width=\columnwidth]{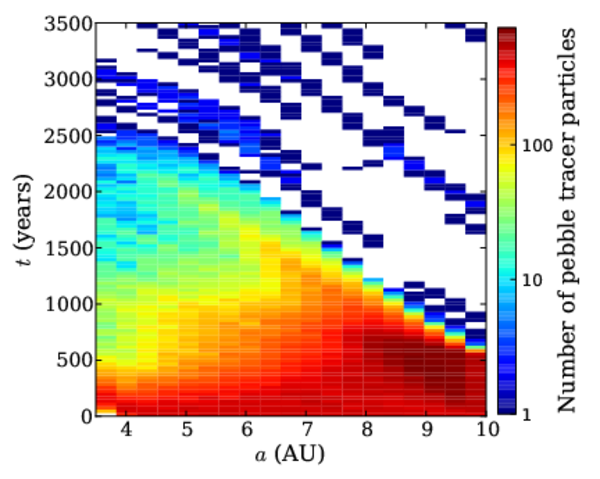}
	\caption{Number of pebble tracer particles as a function of location and time in the fiducial simulation.  Pebbles are added to the outer regions in order to keep the surface density in the outer regions approximately constant for the first 600 yr.  By 3000 yr the pebbles have largely been accreted or drifted out of the simulation.
	\label{fig:surfpebble_fid}
	}
\end{figure}

\begin{figure} 
\includegraphics[width=\columnwidth]{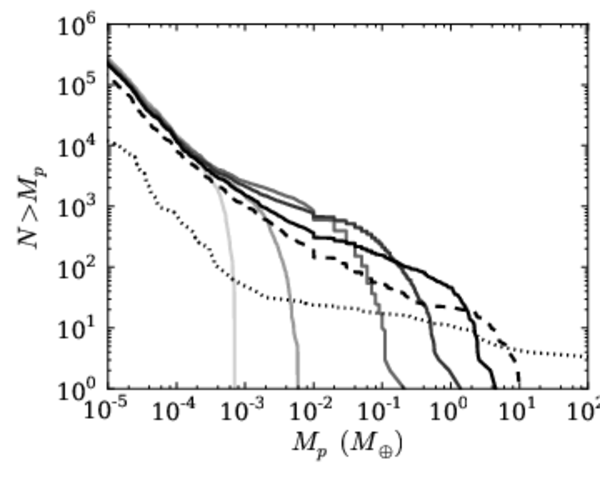}
	\caption{Cumulative mass distribution of planetesimals/planets at different times during the fiducial simulation (described in the text).  The curves show the initial distribution and at 90, 300, 600, and 3000 yr (light to dark solid curves, respectively) as well as at $10^6$ and $5\times10^6$ yr (dashed and dotted, respectively).
	\label{fig:fiducial}
	}
\end{figure}

\begin{figure} 
	\includegraphics[width=\columnwidth]{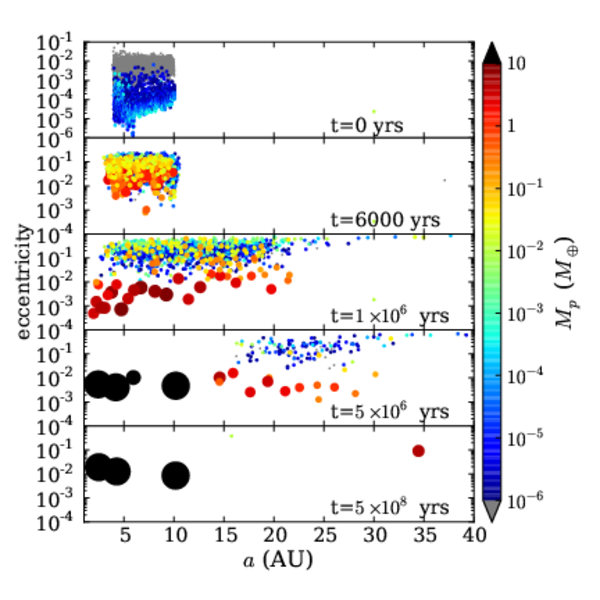}
	\caption{Three snapshots of the eccentricity and semimajor axis of \LIPAD\ particles at different times in the fiducial simulation.
	The pebbles are indicated by small gray points.  Planetesimals and planets are indicated by colored circles; if planets are larger than $10^{-2}$, then the size of the circle is roughly proportional to the planet's radius.
	\label{fig:fiducial_ae}
	}
\end{figure}

\begin{figure} 
	\includegraphics[width=\columnwidth]{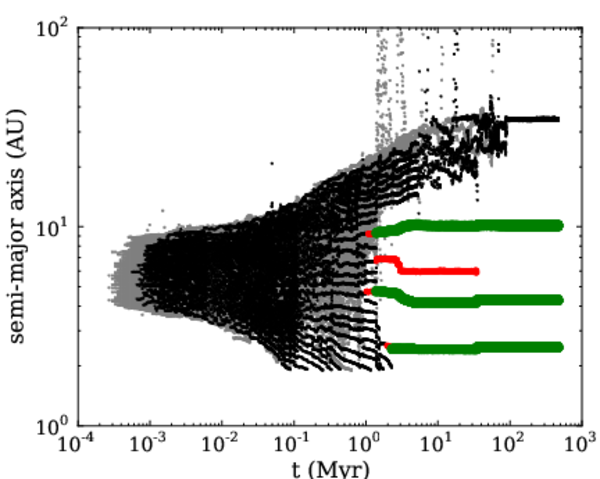}
	\caption{Semimajor axis of particles as a function of time.  Planets are color-coded by mass. Those $>100 M_\oplus$, between 10 and 100 $M_\oplus$, between 1 and 10 $M_\oplus$, and between 0.1 and 1 $M_\oplus$ are indicated by large green points, medium red points, small black points, and gray points, respectively.
	\label{fig:fiducial_a}
	}
\end{figure}

As expected, pebble accretion works extremely well when one has an isolated planetesimal sweeping up pebbles.  However, in this paper we are concerned with a population of planetesimals.  Therefore, for our fiducial model we begin with an initial population of planetesimals between 4 and 10 AU.  We draw these planetesimals from a distribution of $dN/ds \propto s^{-4.5}$ with a minimum planetesimal radius of 100 km ($\sim 10^{-6} M_\oplus$), a maximum size of 1000 km ($\sim 10^{-3} M_\oplus$), and a total planetesimal mass of 20 $M_\oplus$.  For code stability reasons, we also place a single $1.2\times 10^{-6} M_\oplus$ mass planetesimal on a low-eccentricity orbit at 30 AU.  While this particle was intended to not interact with the simulation, it turns out to provide an interesting proxy for the retention of a cold Kuiper Belt and we will return to it later.
We use the same disk model as in our test case, assuming a surface density for the gas disk of $\Sigma=4500 r_{\rm AU}^{-1}$ and that a gas disk has a scale height of $h=0.035 r_{\rm AU}^{5/4} {\rm AU}$.  For these simulations we assume that the gas disk remains constant for the lifetime of the simulation.

Following the results of \citet{Johansen.etal.2009}, we place 20\% of the mass in planetesimals, and the remaining $80 M_\oplus$ is in pebbles.  The pebbles have a size distribution of $dN/dm \propto m^{-3}$ from 0.5 to 5 m.
Pebbles are chosen to have Stoke's numbers near unity, ranging from around $\tau=0.25$ to 25 in the inner regions and from $\tau=0.3$ to 10 in the outer regions of the disk.
Pebbles are given random turbulent kicks consistent with the value that would keep the initial midplane pebble density equal to the gas density.  This is similar to a disk with a \citet{Shakura.Sunyaev.1973} disk with $\alpha=5\times 10^{-3}$.  
We generate new pebbles at 10 AU and allow them to drift inward owing to the aerodynamic drag at a rate of $M_{\rm drift} = 0.13 M_\oplus~{\rm yr}^{-1}$, a rate that maintains the surface density of pebbles in the outer regions.  The surface density of pebble tracer particles is shown in Figure \ref{fig:surfpebble_fid}.
For reference, in Table \ref{tab:runs} we show the parameters for this fiducial run and for the subsequent runs in this paper.

\begin{deluxetable*}{lrrrrrcccccc}
	\tabletypesize{\footnotesize}
	\tablecolumns{11}
	\tablewidth{0pt}
	\tablecaption{ The parameters for the models presented in this paper.
     \label{tab:runs}}
	\tablehead{
	\colhead{}                    & 
	\colhead{$\Sigma_0$}          & 
	\colhead{$h_0$}               & 
	\colhead{}                    & 
	\colhead{}                    & 
	\colhead{$\delta$ t}          & 
	\colhead{}                    & 
	\colhead{pebbles}             & 
	\colhead{pebbles}             &
	\colhead{}                    & 
	\colhead{}                   \\
	\colhead{Run}                 & 
	\colhead{($\rm {g cm^{-2}}$)} & 
	\colhead{(AU)}                &
	\colhead{$q$}                 & 
	\colhead{$p$}                 & 
	\colhead{(years)}             & 
	\colhead{pebble size}         & 
	\colhead{intermixed}          & 
	\colhead{from outside}        & 
	\colhead{evaporation}         & 
	\colhead{planet trap}         }
\startdata
Fiducial&4500 & 0.035&1.25 &-1.0 &0.07& 0.5--5m &\checkmark &\checkmark &&\\	
Fiducial -- Run 2&4500 & 0.035&1.25& -1.0&0.07&0.5--5m &\checkmark &\checkmark &&\\
Fiducial Shielding&4500 &0.035 &1.25 &-1.0 &0.07& 0.5--5m& &\checkmark &&\\
Large $\eta$ & 17\,400 & 0.035 &1.35 &-1.9 &0.07& 0.5--5m &\checkmark &\checkmark &&\\
Small $\eta$ & 1150 & 0.035 & 1.15 & -0.1 &0.07& 0.5--5m &\checkmark &\checkmark &&\\
Small Pebbles&4500 &0.035 &1.25 &-1.0 &0.01 & 0.2--0.5m &\checkmark &\checkmark &&\\
Large Pebbles&4500 &0.035 &1.25 &-1.0 &0.2& 5--10m &\checkmark &\checkmark &&\\
Large Shielding&4500 &0.035 &1.25 &-1.0 &0.2& 5--10m & &\checkmark &&\\
Snow line &4500&0.035&1.25 &-1.0 &0.07& 0.5--5m &\checkmark & & \checkmark &\\
Planet Trap &4500&0.035&-0.5& -1.0 &0.07& 0.5--5m &\checkmark & & \checkmark &\checkmark\\
\enddata
\vspace{-0.2cm}
\end{deluxetable*}

We continue generating pebbles for 500 yr, leading to 65$M_\oplus$ of pebbles entering the domain.
This rather high flux illustrates one of the peculiarities in this model. Because pebbles drift inward very rapidly, in order to maintain the pebble surface density at the initial level we must supply them at a high rate.
This is a rather high flux of pebbles, and in an actual disk the mass flux of pebbles might be slower and spread out over a longer period of time.  However, this would simply slow down the accretion of pebbles relative to the other growth processes going on.  Given the fact that we are interested in testing whether pebble accretion can be the dominant mechanism we choose to use these large pebble flux rates. 

In Figure \ref{fig:fiducial} we show the initial cumulative distribution of particles in light gray and the distribution after 90, 300, 600, 3000, $10^6$, and $5\times10^6$ yr of evolution in subsequently darker curves (with the final two curves dashed and dotted, respectively).  
One can see that even after a very short amount of time the largest planetesimals have grown quite large.  As time progresses, the sharp cutoff in mass at the largest size softens and shallows as a few of the planetesimals are lucky and accrete more pebbles than their siblings.  However, this runaway growth is not enough to create only a few proto-cores.  Instead, by 600 yr almost 200 planetesimals (now protoplanets) grow to above Mars mass in size and none grow to much above an Earth mass.
By 3000 yr all of the pebbles have been lost and there are almost 50 protoplanets more massive than 1 $M_\oplus$, with the largest being around 4.5 $M_\oplus$.

In the subsequent million years there is some evolution of this profile in which a few of the largest embryos grow to around 10 $M_\oplus$ by eating their siblings, but there are still $\sim30$ planets more massive than Earth. 
Because some of the cores have grown large enough to accrete gas, by 5 Myr three of the cores have become gas giants.  
At 1 Myr there are still eight non-gas planets larger than 1 $M_\oplus$ and another half-dozen objects between Mars and Earth mass.  

Despite the growth of three gas giants, the plethora of Mars- and Earth-mass bodies generated in this fashion provide a significant problem.  The reason for this can be seen by looking at the evolution of the embryos' semimajor axes.  In Figure \ref{fig:fiducial_ae} we show the semimajor axis and eccentricity distribution at five different times during the simulation; the initial state and the states after 6000, $10^6$, $5 \times 10^6$, and $5\times 10^8$ yr.  Early on, the embryos grew extremely rapidly by sweeping up the pebbles.  Accretion was so fast that they did not have time to dynamically interact with one another.  In the subsequent million years a few of these Earth-mass protoplanets collide and grow, however the vast majority of them scatter and spread instead.  This means that even after the few lucky cores that are able to accrete gas form, there are many Earth-mass planets in orbits that are farther from the Sun.  
We continue the integration of these remaining bodies for a few hundred Myr in a gas-free environment.
Eventually these planets become gravitationally unstable, crossing orbits.  Some collide and merge, while others are ejected from the system.
The initial low-mass planetesimal at 30 AU is completely removed, confirming that this process has catastrophically disrupted the region where today we would expect the Kuiper Belt to have formed.
This scenario is clearly incompatible with our own solar system.  

Additionally, to speed up the computations, we removed objects that had semimajor axis below 2 AU.  
To more clearly see what has been lost by this assumption, in Figure \ref{fig:fiducial_a} we show the semimajor axis evolution of the protoplanets within the computation.  At various time steps we indicate Mars-mass, Earth-mass, 10 $M_\oplus$, and 100 $M_\oplus$ planets with progressively larger points.
By looking at the semimajor axis evolution of the protoplanets we can see that there were eight objects greater than 1 $M_\oplus$ that were lost because they crossed into the inner solar system.  
We would expect many of these (water-rich) bodies to have survived in this region, disrupting any formation of terrestrial planets.

\begin{figure} 
	\includegraphics[width=\columnwidth]{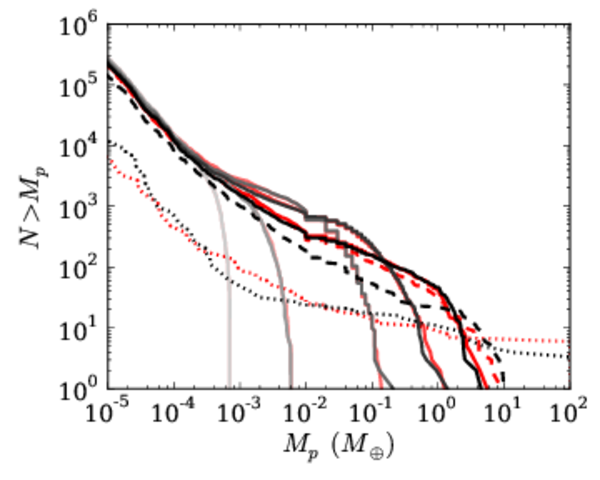}
	\caption{Cumulative mass distribution for a second run in the fiducial disk.  The curves are for the same times as in Figure \ref{fig:fiducial}, and, for ease of comparison, they are overplotted in black on top of those from the original run shown in red.
	\label{fig:fiducial2}
	}
\end{figure}

\begin{figure} 
	\includegraphics[width=\columnwidth]{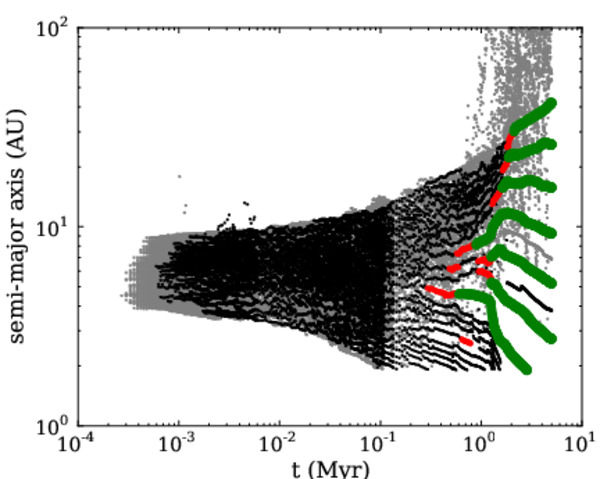}
	\caption{Semimajor axis as a function time of particles in Run 2 in the fiducial disk.  The symbols are the same as in Figure \ref{fig:fiducial_a}.
	\label{fig:fiducial2_a}
	}
\end{figure}

\begin{figure} 
	\includegraphics[width=\columnwidth]{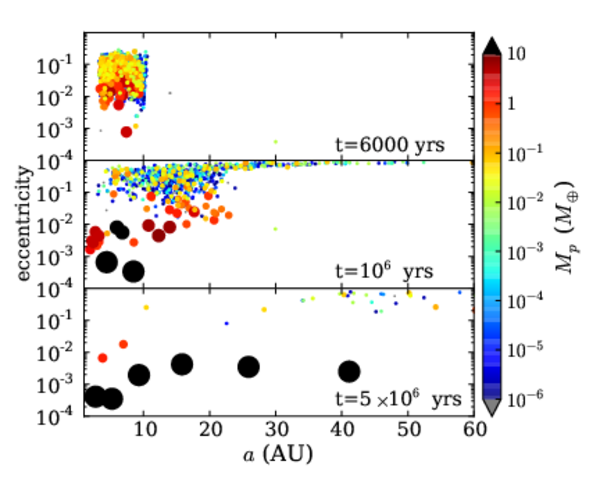}
	\caption{Snapshot of the eccentricity and semimajor axis of particles at three different times in run 2 of the fiducial disk. The symbols are the same as in Figure \ref{fig:fiducial_ae}.
	\label{fig:fiducial2_ae}
	}
\end{figure}

Furthermore, we may be overly optimistic in identifying these final objects as gas giants.  Remember that our model for giant planet formation is simple; once the core mass is greater than the critical core mass as a function of its current mass accretion rate we allow it to accrete gas.   Subsequent impacts cannot reduce the planet's gaseous envelopes.  This means that despite the fact that the giant planets are impacted with high-velocity collisions with massive $> M_\oplus$ embryos, we do not allow erosion of the atmosphere as would be expected \citep{Broeg.Benz.2012}.  

Of course, the fact that the oligarchs reached around an Earth mass was a function of the amount of pebbles we injected into the simulations.  If we had kept adding pebbles to the disk, then the planetesimals would eventually reach 10 $M_\oplus$ without relying on mutual collisions.  However, the relative growth rates would be the same.  With no clear protoplanet running away, instead we would have perhaps one hundred 10 $M_\oplus$ cores competing to become gas giant planets, clearly unlike anything within our own solar system.  On the other hand, if we had added fewer pebbles then we would have been left with a population of lower mass embryos.  This would look very similar to the traditional planet formation scenario except for the fact that most of the mass in the system is in the large oligarchs, so the growth timescale would be even longer than the classical case.

To test the robustness of the result of our fiducial run, we ran another calculation with the same disk and pebble properties, but with different randomly placed planetesimals as our initial conditions.  In this second run we saw that the early growth of the planetesimals was very similar to the first calculation (Figure \ref{fig:fiducial2}).
However, the giant planets formed a little closer to each other and scattered each other more strongly.  This caused one giant planet to move outward, causing the collisions of other protoplanets (Figure \ref{fig:fiducial2_a}).  Because we keep the gas surface density constant throughout this simulation, even these later-forming massive cores can accrete gas and become giant planets.  
In the end this left a system of six relatively low eccentricity Jupiter-mass planets (Figure \ref{fig:fiducial2_ae}).  However, there still are several Earth-mass objects that drifted inward out of the computational domain, and the Mars-mass bodies were scattered throughout the outer solar system.  So while this system may look tantalizingly like some known exoplanet systems \citep[such as HR 8799,][]{Marois.etal.2008} it is clearly not what happened in our own solar system.

\section{Semianalytic Growth Rates} \label{sec:analytics}
In the fiducial runs it appears that the growth of the planetesimals resembles the standard so-called oligarchic regime in the planet formation process \citep{Kokubo.Ida.1998,Kokubo.Ida.2000}.  In this regime the mass doubling time for the largest object becomes long compared to smaller objects, allowing the smaller ones to grow relatively faster and ``catch up'' to the larger objects.  This is in contrast to the ``runaway'' regime in which a slightly larger object has a shorter mass doubling time, allowing it to grow significantly (and increasingly) faster than its smaller neighbors. 

To understand the observed evolution better, we take a closer look at the growth rate of planetesimals due to pebble accretion.  Let us define
\begin{equation}
	\xi \equiv \frac{d\ln\dot M_p}{d\ln M_p}.
	\label{eq:p}
\end{equation}
If $\xi > 1$, then we are in a regime in which the doubling time for the largest (proto-)planets is the shortest, allowing the biggest object to run away from its smaller neighbors.  This causes the cumulative size distribution to become shallower. However, if $\xi < 1$ then the largest objects have mass doubling times that are longer than the smaller ones.  This means that the smaller objects catch up to the larger ones, causing the size distribution to steepen.  At $\xi = 1$ the slope of the size distribution remains unchanged.

\begin{figure} 
\includegraphics[width=\columnwidth]{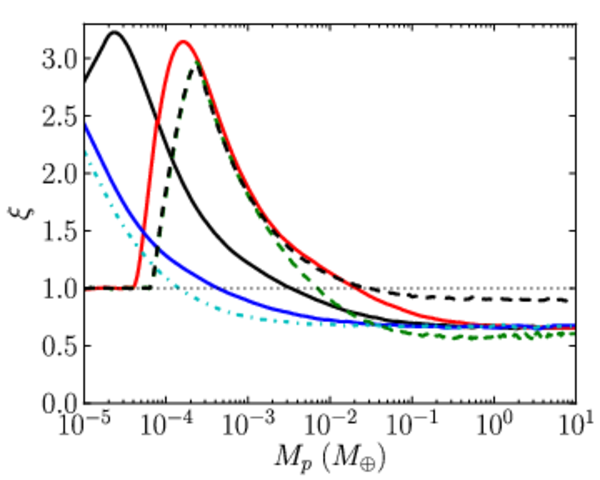}
	\caption{Growth exponent, $\xi$, as a function of planet mass for different assumptions about the disk and pebble parameters.
The fiducial model is shown in solid black, while in red 
we show the results for a more strongly pressure-supported disk (large $\eta$ in Table \ref{tab:runs}), and in blue the results for a shallower disk model (Small $\eta$).
The dashed green curve shows the results for the small particles if turbulence is decreased so that the particle midplane density is the same as the fiducial model.  
The dashed black curve show the result for the fiducial disk model with small pebbles such that it remains in the 3D accretion regime to high masses.
The cyan dash-dotted curve shows the results for large ``pebbles''.
	\label{fig:exponent}}
\end{figure}

By investigating Equation (\ref{eq:Mdot}), we can semianalytically calculate $\xi$.  In Figure~\ref{fig:exponent} we show $\xi$ as a function of planet mass for the pebble sizes and disk properties in our fiducial model.
In these plots we include not only the pebble accretion cross section described in Section \ref{sec:review} but also gravitational focusing as described as the hyperbolic regime in \citet{Ormel.Klahr.2010} (their Equation (28)).  This is only relevant for the smallest planetesimals.
We can see that for small planetesimals $\xi>1$, indicating runaway-type growth.  However, at around $M_p = 3\times 10^{-3} M_\oplus$ the curve dips below unity, indicating that the growth has become oligarchic-like.  This is consistent with Figure~\ref{fig:fiducial}, in which we see an initial shallowing of the size distribution early in the simulation, but as the pebbles continue to accrete, there is a steepening of the size distribution at high masses as a significant number of objects grow together, as oligarchs.

One way to avoid the problem of having too many Mars- and Earth-mass objects at the time of giant planet formation is to maintain runaway accretion all the way to $10 M_\oplus$.  
To investigate whether this is possible, we can approximate the growth rate as $\dot M = \sigma\vrel\rho_{\rm peb}$, where $\sigma$ is the effective capture cross section and $\rho_{\rm peb}$ is the density of pebbles at the midplane.  Therefore, $\xi$ is determined by a simple combination of the two effects,
\begin{equation}
	\xi = \frac{d\ln\sigma}{d\ln M_p} + \frac{d\ln\vrel}{d\ln M_p}. \nonumber
\end{equation}

Let us first look at the capture cross section.  
If the scale height of the pebbles ($h_p$) is much larger than the capture radius ($\Rcap$), then accretion is fully three-dimensional (3D) and the effective capture cross section is well approximated by $\sigma \approx \pi \Rcap^2$.
However, if $\Rcap$ is much larger than $h_p$, then accretion is effectively two-dimensional (2D) and $\sigma\approx 4 \Rcap h_p$.
This means that
\begin{equation}
	\frac{d \ln \sigma}{d \ln M_p} = f_1 \frac{d\ln\Rcap}{d\ln M_p} \nonumber
\end{equation}
where $f_1=1$ in the 2D case and $f_1=2$ in the 3D accretion.

To understand the behavior of $R_{\rm cap}$ as a function of $M_p$ we can look at the two extreme solutions to Equation (\ref{eq:Rcap}), the shear-dominated regime (Equation (\ref{eq:R_shear})) or the head-wind-dominated regime (Equation (\ref{eq:R_headwind})).  In these two extremes, 
\begin{equation}
	\frac{\partial \ln \Rcap}{\partial \ln M_p} = f_2 + \gamma \left(\frac{M_*\eta^3\Theta^3\tau}{4 M_p}\right)^\gamma,
	\label{eq:dRcapdM}
\end{equation}
where in the shear-dominated regime $f_2=1/3$ and in the drift regime $f_2=1/2$.
Since we are interested in the growth rates of larger planets, so long as we limit ourselves to large planets in which
\begin{equation}
M_p \gg M_*\eta^3\Theta^3\tau,
\label{eq:M_trans}
\end{equation}
then the second term in Equation (\ref{eq:dRcapdM}) is negligible. 
Additionally, so long as $\tau \lesssim 1$, this is the same condition for being in the shear-dominated regime.  Therefore, when looking at large planets, we can conclude that $d\ln \Rcap/d\ln M \approx 1/3$.

To understand the behavior of $v_{\rm rel}$ as a function of $M_p$, we can look at the relative velocity from Equation (\ref{eq:vrel}) and take $x=R_{\rm cap}$. This leads to
\begin{equation}
 \frac{d\ln\vrel}{d\ln M_p}  = \left(1+\frac{a\eta \Theta}{\Rcap}\right)^{-1} \frac{d \ln\Rcap}{d\ln M_p}. \nonumber
\end{equation}
Therefore,
\begin{equation}
\frac{d\ln\dot M_p}{d\ln M_p} = \frac{d\ln\Rcap}{d\ln M_p} \left[f_1+\left(1+\frac{a\eta\Theta}{\Rcap}\right)^{-1}\right]. \nonumber
\end{equation}
The bracketed part of this equation approaches $f_1+1$ for small planets (when $\Rcap/a \ll 2 \eta \Theta/3$), and in the other extreme it approaches $f_1$.
Therefore we can easily conclude that, for a monodisperse pebble distribution (and probably for other distributions as well), at high masses (so that $d\ln\Rcap/d\ln M=1/3$) the only way to avoid oligarchic growth is to be in the 3D accretion regime ($f_1=3$). 
It is extremely hard to keep $\tau=1$ pebbles in the 3D accretion regime at high masses.  For example, to keep the particle scale height of $\tau=1$ pebbles larger than the accretion capture radius of a 1 $M_\oplus$ planet in a disk with an $h/r = 0.05$, there must be $\alpha \simeq 0.3$ \citep[following][]{Youdin.Lithwick.2007}.  So for more realistic turbulence high excitation requires small pebbles.
And even in those most favorable cases, where the pebbles are lofted high enough to keep the growth in the 3D regime, we only expect $\xi=1$.  This is a marginal case, not a strong runaway.

By using Equation \ref{eq:Mdot} we can calculate the $\xi$ parameter as a function of $M_p$ for different pebble sizes and disk structure parameters.
In Figure \ref{fig:exponent} we compare the fiducial model to other disk models (whose disk parameters are shown in Table \ref{tab:runs}).
In red we show the results for a strongly pressure-supported disk (large $\eta$ in Table \ref{tab:runs}), resulting in a faster drift speed with $\eta = 6.4\times 10^{-3}$.
Note that at the lowest masses the physical radius of the planetesimal determines the cross section.  In blue we show the results for a flatter disk model with $\eta = 1.9\times 10^{-3}$ (small $\eta$) .

\begin{figure} 
	\includegraphics[width=\columnwidth]{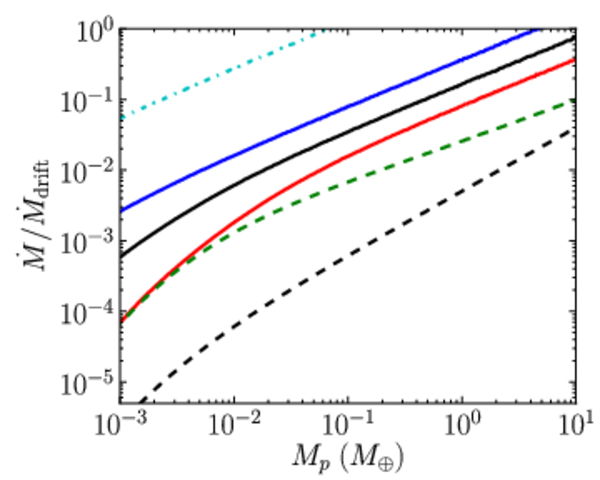}
\caption{Approximation of the probability that a pebble is captured by a planet or planetesimal of mass $M_p$ for the same disk and pebble parameters as shown in Figure \ref{fig:exponent}. 
	\label{fig:Peff}}
\end{figure}

We cannot avoid the ``oligarchic dilemma'' through moderate reasonable changes to the disk parameters.
In order to maintain runaway accretion until an embryo reaches 10 $M_\oplus$, for particles near $\tau\approx 1$, we must have $\eta>3\times 10^{-2}$.  Not only is this $\eta$ value an order of magnitude higher than expected in typical models (such as the MMSN), but in a disk that is so strongly pressure supported the pebbles would drift inward an order of magnitude faster, strongly decreasing the capture probability (as discussed subsequently and shown in Figure \ref{fig:Peff}).

\begin{figure} 
\includegraphics[width=\columnwidth]{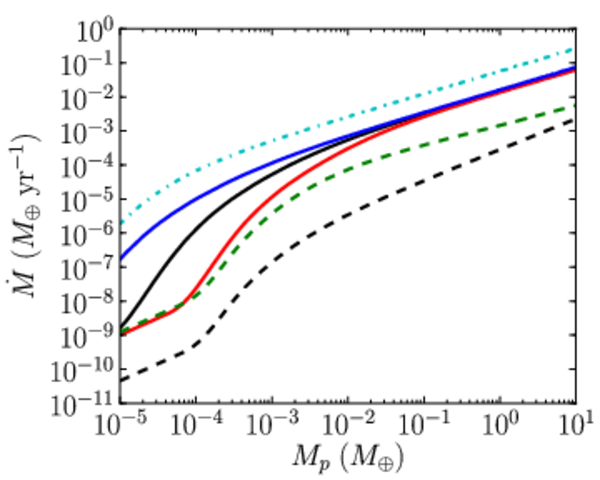}
\caption{Pebble mass accretion rate onto planets or planetesimals of mass $M_p$ for the same disk and pebble models as shown in Figure \ref{fig:exponent}. 
	\label{fig:Mdot}}
\end{figure}

The only way to change the high-mass asymptote of the $\xi$ curve is to keep the growth in the 3D regime.  
To demonstrate this, in Figure \ref{fig:exponent} we compare two models with different turbulent strengths such that one is in the 2D regime at large masses, while the other is in the 3D regime.
The dashed green curve shows the results for a disk with small pebbles when the disk turbulence is set so that the particle scale height (and thus midplane density) is the same as in the  fiducial model.  
Compare this to the dashed black curve, which shows the result for the same fiducial disk model and small pebbles (parameters in Table \ref{tab:runs}) with a higher pebble scale height such that protoplanets remain in the 3D accretion regime to high masses.
The two curves diverge at high mass.
The fact that $\xi$ is slightly less than unity at high mass is due to the small spread in pebble sizes.  However, even if $\xi$ had been identically one (as we would expect for a monodisperse pebble population), this would only preserve the size distribution.  We cannot preserve runaway growth to high masses, and with a spread of pebble sizes, we cannot even reach the marginal case.

Additionally, in Figure \ref{fig:Mdot} we show that this nearly marginal growth rate comes at a price.  The mass accretion rate (for the same pebble surface density) is more than an order of magnitude lower for the excited small pebbles model compared to the fiducial model.
In order to not fall deeply into oligarchic growth, the accretion rate is necessarily low. 
This low accretion rate is problematic because pebbles have a limited life span in the disk. The particles drift inward at a rate of
\begin{equation} 
	\dot M_{\rm drift}= 2\pi \int_{s_{\rm min}}^{s_{\rm max}} \frac{d\Sigma}{ds} a v_r ds \nonumber
\end{equation}
where
\begin{equation}
	v_r= 2\frac{\tau}{\tau^2+1}\eta v_K \nonumber
\end{equation}
\citep{Weidenschilling.1977}.
We can roughly estimate the probability that a particle is accreted by looking at the ratio of 
\begin{equation}
	P_{\rm eff} \approx \frac{\dot M_p}{\dot M_{\rm drift}}. \nonumber
\end{equation}
Figure \ref{fig:Peff} shows $P_{\rm eff}$ for the various disk models.
Although large planets (above 1 $M_\oplus$) have fairly substantial capture probabilities, the probability for a pebble to be accreted is quite small for small planetesimals.  This means that it takes quite a lot of pebbles to initially grow a planetesimal from, say, $10^{-2} M_\oplus$ to $1 M_\oplus$.
This is not insurmountable.  However, if one looks at a disk with small excited pebbles, then the problem becomes much more acute.
Several orders of magnitude more mass must have drifted past the planetesimal without being accreted in order to have the same growth rate as the larger, more settled pebbles.
This means that to even just get that almost marginal growth rate, we have to sacrifice the efficiency that makes pebble accretion attractive.

\section{Basic Model Variations} \label{sec:eta_size}
\subsection{Disk Structure} \label{sec:eta}
In the previous section we can see that simple analytic estimates suggest that the {\it oligarchic growth problem} described above cannot be solved by changing the structure of the disk.  
However, to confirm the validity of the analytic approximations and test whether other physics in the complete \LIPAD\ simulations could possibly lead to a solution, we run a series of simulations where we vary the disk structure.

\begin{figure} 
	\includegraphics[width=\columnwidth]{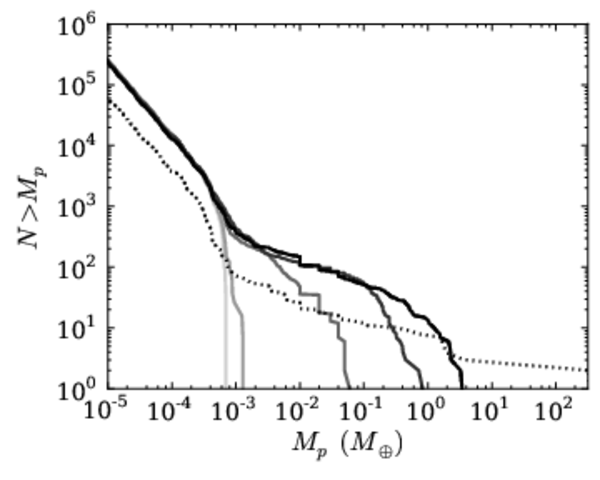}
	\caption{Cumulative distribution in a disk with a large $\eta$.  The curves are the same as for Figure \ref{fig:fiducial}.
	\label{fig:largeeta}
	}
\end{figure}

In Figure \ref{fig:largeeta} we show the evolution of the size distribution for a disk with a larger $\eta$ value ($q=1.15$, $p=-1.9$, which leads to $\eta=6.8\times10^{-3}$ at 5 AU).
We have adjusted the disk profile so that the midplane density is the same at 4.5 AU as our fiducial model. Note that this leads to an extremely high surface density at 1 AU ($\Sigma_0 = 17~400 {\rm g~cm}^{-2}$).  Thus, this disk is probably not physical and should be considered an extreme case.

\begin{figure} 
	\includegraphics[width=\columnwidth]{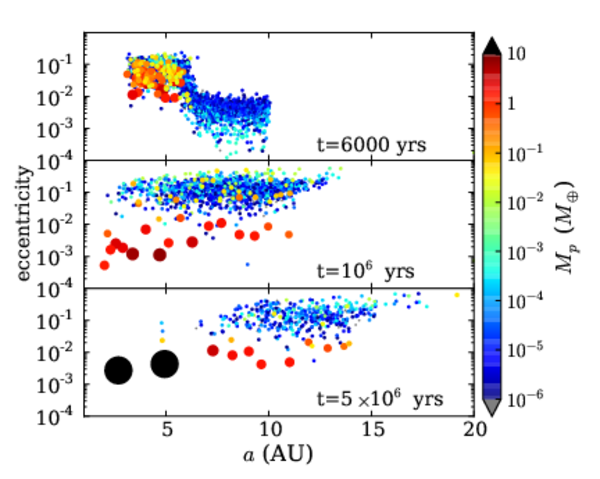}
	\caption{Snapshots in the evolution of a disk with a larger $\eta$.  Symbols are the same as in Figure \ref{fig:fiducial_ae}. 
	\label{fig:largeeta_ae}
	}
\end{figure}

Compared to the fiducial model, we see that, during the few thousand years in which the pebbles persist, there is a notably shallower slope in the size distribution between $10^{-3}$ and $10^{-1} M_\oplus$.  This indicates that the planetesimals had a stronger runaway phase that extended to larger planet masses.  However the effect is still not strong enough to avoid producing a population of Earth-mass planets that are not incorporated into giant planets (as seen in Figure \ref{fig:largeeta_ae}).

\begin{figure} 
	\includegraphics[width=\columnwidth]{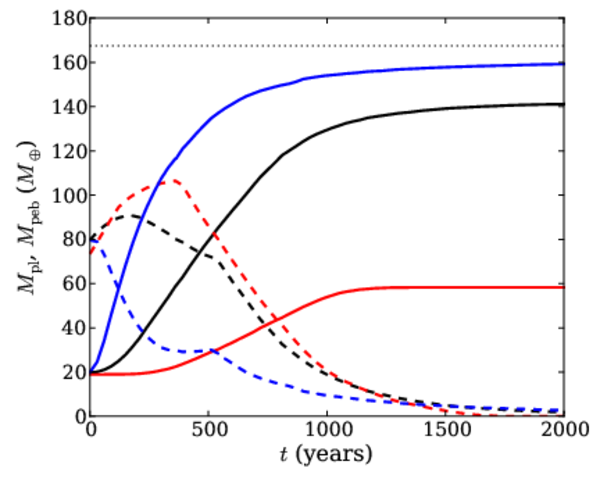}
	\caption{Amount of material in the disk as a function of time that is in planetesimals/planets (solid curves) compared to the amount in pebbles (dashed curves).  The red, black, and blue curves refer to the large-$\eta$ disk, the fiducial disk, and the disk with a small $\eta$ value, respectively.  We place a dotted line to mark the total amount of solids that enter into the simulations.
	\label{fig:eff_eta}
	}
\end{figure}

Despite the fact that the total amount of solids that enter the computational domain in the above run is the same as in the fiducial case, the total amount of solids retained in the domain after the pebbles disappear is significantly smaller.  
As we show in Figure \ref{fig:eff_eta}, the total amount of solids converted into planetesimals varies significantly as a function of the $\eta$ parameter in the disk.
In this large $\eta$ model (red curve) only around 40 $M_\oplus$ of pebbles (less than 30\% of the pebbles) are accreted by the planetesimals. 
This compares with the 80\% retention in the fiducial model.
This agrees with what we would expect from the analytic estimates in Figure \ref{fig:Peff}.

\begin{figure} 
	\includegraphics[width=\columnwidth]{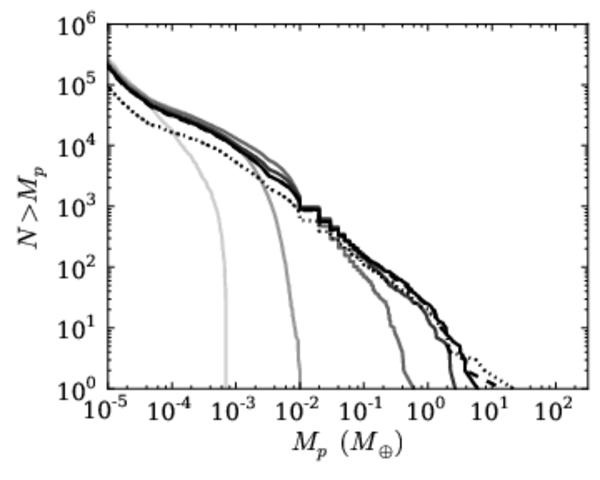}
	\caption{Cumulative distribution in a disk with a small $\eta$.  The curves are the same as for Figure \ref{fig:fiducial}.
	\label{fig:smalleta}
	}
\end{figure}

\begin{figure} 
	\includegraphics[width=\columnwidth]{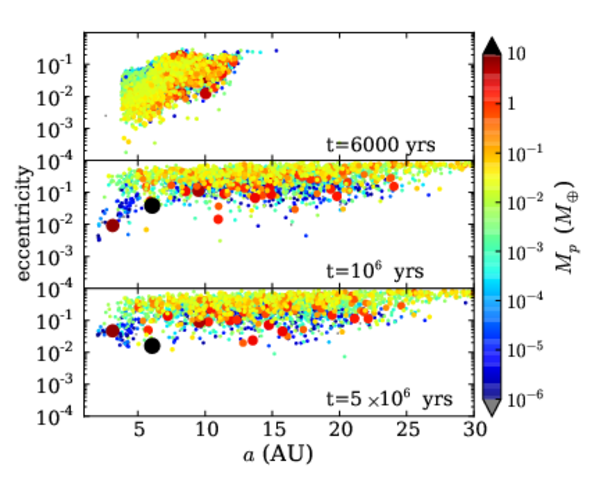}
	\caption{Snapshots in the evolution of a disk with a smaller $\eta$.  Symbols are the same as in Figure \ref{fig:fiducial_ae}.
	\label{fig:smalleta_ae}
	}
\end{figure}

When we look at a less pressure-supported disk ($\Sigma_0=1150{\rm g~cm}^{-2}$, $h_0=0.035 {\rm AU}$, $q= 1.15$, $p= -0.1$, $\eta= 2.0\times10^{-3}$ at 5 AU) with a slower particle drift speed, we see that, except for the very smallest planetesimals, there is little evidence of runaway growth (Figure \ref{fig:smalleta}).  
Indeed, in this model no giant planets are able to form within 5 Myr despite the formation of a couple of 10 $M_\oplus$ cores.  
Recall that in order to accrete gas the core needs to become larger than the critical core mass, but that the value of the critical core mass is a function of the accretion rate of solids \citep{Mizuno.1980,Stevenson.1982}.  If the solid accretion rate is too high, then the core's atmosphere is too hot to accrete more gas.  In this run, these $10 M_\oplus$ cores are actually still subcritical owing to the high solid accretion rates, so they do not become gas giants.
In Figure \ref{fig:smalleta_ae} we show three snapshots of the disk, demonstrating the large number of excited planetary-mass objects.
This shows an additional constraint, not only do we need large embryos, but we also need to shut off solid accretion.

As can be seen in Figure \ref{fig:eff_eta}, in the low-$\eta$ model (the blue curve) nearly 95\% of the pebbles that enter into the domain are accreted by the planetesimals.
This high degree of retention is what we expected; the timescale to accrete material is much shorter than the timescale for it to drift, even when the planetesimals are quite small. 
Therefore, the large- and small-$\eta$ runs confirm that we can get either high retention and many oligarchs, or low retention and (even for an unrealistic disk) an only marginally more favorable size distribution.

All together, the results are as we would expect from the simple analytic model. We conclude that changes to the disk structure cannot, on their own, produce results consistent with our own solar system.

\subsection{Pebble Size} \label{sec:size}
In Section \ref{sec:analytics} we argued that for small, more excited particles (i.e.,~particles that are stirred by outside turbulence rather than self-stirring) it is possible to maintain a growth exponent close to one even at high masses.
This is due to the fact that the scale height of the particle remains large compared to the capture cross section of these small particles, keeping the planet in the 3D accretion regime.

\begin{figure} 
	\includegraphics[width=\columnwidth]{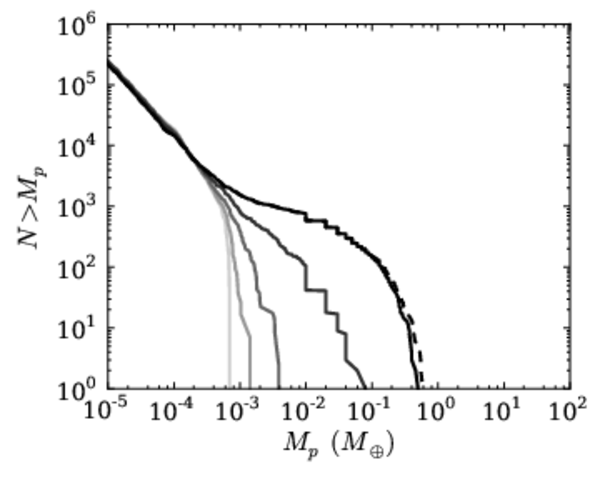}
	\caption{Cumulative distribution in the fiducial disk with small pebbles.  The curves are 0, 90, 300, 600, and 3000 yr (light to dark curves) and at $1.1 \times 10^5$ yr (dashed curve).  Note that this differs from those in Figure \ref{fig:fiducial} because we terminated the run early owing to the fact that no objects have reached $1 M_\oplus$.
	\label{fig:small}
	}
\end{figure}

\begin{figure}
	\includegraphics[width=\columnwidth]{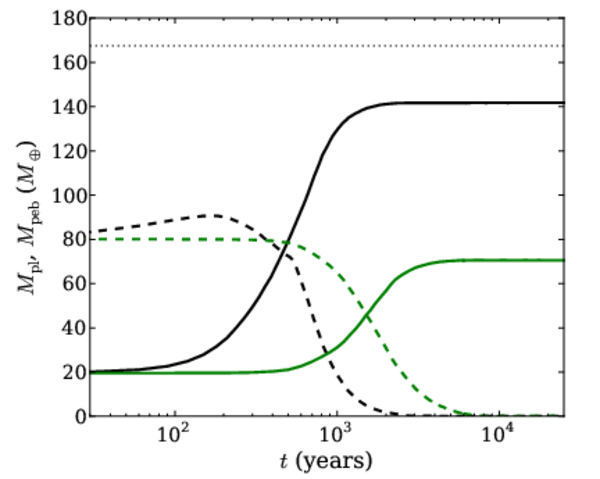}
	\caption{Similar to Figure \ref{fig:eff_eta} but comparing the pebbles in the fiducial run (black) to those in the small-pebble run (green).  Note that time is shown on a log scale because the pebble drift timescales are very different in the two runs.
	\label{fig:eff_small}}
\end{figure}

To test this effect, we performed a small pebble run where we include pebbles with radii of 20--50 cm and placed them in our fiducial disk, but forced the initial midplane pebble-to-gas ratio to be 0.03 instead of the usual 1.0.  This lower gas-to-dust ratio is equivalent to keeping the turbulent $\alpha$ parameter the same as in our fiducial run.  However, we note that this higher level of turbulence is not directly consistent with efficient planetesimal formation via the streaming instability.
In Figure \ref{fig:small} we show the cumulative distribution of planetesimals at different times.  
The pebbles have all disappeared by 6000 yr (either by being accreted or by drifting inward), and no planetesimal has grown to more than an Earth mass.  In the next $10^5$ yr the size distribution does not change notably.
The slight steepening of the high mass slope between 600 and 3000 yr indicates that we have still transitioned into the oligarchic growth regime even in this more favorable case.
The total amount of growth was much smaller in this run as a result of the low accretion efficiency.
In Figure \ref{fig:eff_small} we show that a much smaller fraction of the pebbles are accreted in this simulation than in the fiducial one.

\section{Initial planetesimal population} \label{sec:planetesimals}
Another, perhaps even larger, unknown in these models is the mass and size distribution of the initial planetesimal population.
In the previously presented simulations we have assumed that planetesimals form within a size range of 100--1000 km, have a slope of $dN/ds \propto s^{-q_{\rm pl}}$ (where $q_{\rm pl} = 4.5$), and have a total mass of 20 $M_\oplus$.  
This size distribution was motivated by studies that suggest that the high-mass end of the existing small-body reservoirs in our solar system may be indicative of the  initial populations \citep{Nesvorny.etal.2010,Morbidelli.etal.2009}.
Observationally, $q_{\rm pl}\sim 4.5$ is roughly consistent with the present-day slope for objects larger than 50 km \citep{Trujillo.etal.2001, Bernstein.etal.2004,Fuentes.Holman.2008, Fraser.etal.2008,Fraser.etal.2010a}.
The total initial amount of solids in our computational domain was chosen to be 100 $M_\oplus$, a round number roughly comparable to the total amount of solids currently in the giant planets in our solar system \citep{Hayashi.1981}.
We set the percentage of mass in planetesimals to be 20\%, which is roughly consistent with streaming instability calculations \citep{Johansen.etal.2009}.
This mass is also comparable to the mass needed for the postulated so-called Nice disk, the planetesimal disk that was necessary to move the giant planets into their current configurations \citep{Gomes.etal.2005,Tsiganis.etal.2005}.

Despite these motivations, very little is actually known about the size distribution of planetesimals expected in the formation process.
While theoretical models have shown that it is possible to form planetesimals even as large as $10^{-3} M_\oplus$ via the streaming instability in only a few local orbital periods \citep{Johansen.etal.2007,Chiang.Youdin.2010}, these models do not yet have predictions of the expected size distribution of planetesimals.
If planetesimals form with a significantly different size distribution than assumed in this study, one could get very different results.
For example, LJ2012 suggested that planetesimals may be produced in such a way that only a few (four or five in our solar system) are large enough to accrete pebbles efficiently, while the rest are too small.  In this case the oligarchic dilemma we have described will never arise because there are not enough planetesimals available.  
However, as we address in detail in the remainder of this section, the LJ2012 model requires a very special set of initial conditions, which we believe may be difficult to achieve.

\begin{figure} 
	\includegraphics[width=\columnwidth]{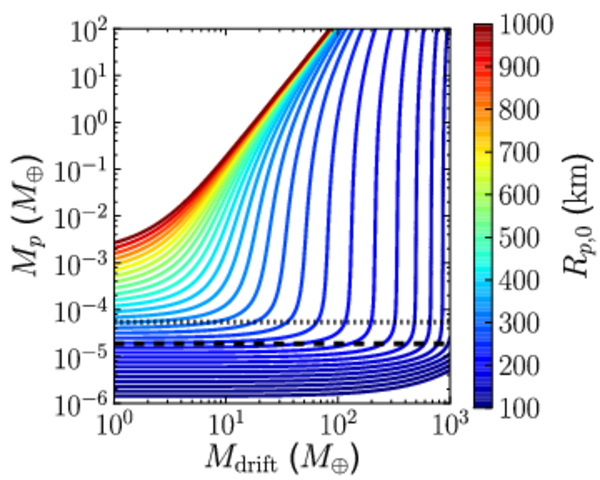}
	\caption{Growth curves for planetesimals of different sizes at 5 AU in the fiducial disk with the fiducial pebble size distribution following Equation (\ref{eq:Mdot}), as a function of the cumulative amount of pebbles that have drifted past their orbit. 
		The curves are color-coded by the initial planetesimal size.  For reference, our preferred criterion for the critical mass is indicated as a black dashed line (Equation (\ref{eq:Mcrit}), $C=2$), while the $C=1$ curve is indicated by the dotted black line.
	\label{fig:growth_Disk1}
	}
\end{figure}

To aid in our discussion of the initial conditions required by LJ2012, in Figure \ref{fig:growth_Disk1} we show the growth curves for planets in the fiducial disk as a function of amount of material drifting past them (allowing the cores to grow to arbitrarily large sizes).
If we want a 1000 km planetesimal to grow to $10 M_\oplus$, then around $40 M_\oplus$ of pebbles must have drifted by it.  During that time, any nearby planetesimal much larger than 300 km would also have grown significantly. The tendency toward oligarchic growth means that if a planetesimal is large enough to have grown, then it also will grow to approximately the same mass as all of the objects larger than it.
It is important to note that there are no planetesimals with zero growth rates as a planetesimal can always accrete owing to direct impacts, but the timescale for a planetesimal to grow becomes significantly longer as one looks at subsequently smaller initial planetesimals.  

\begin{figure} 
	\includegraphics[width=\columnwidth]{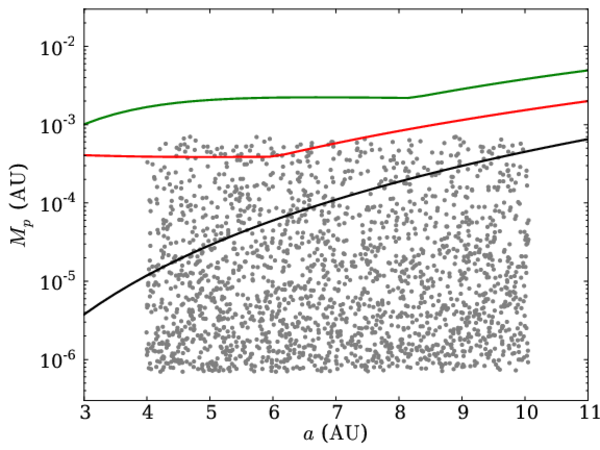}
	\caption{Critical curves (Equation (\ref{eq:Mcrit}), $C=2$) for planet growth due to pebble accretion in the fiducial disk model for pebbles of size 500, 100, and 50 cm (in black, green, and red, respectively).  
	The curves are overlaid on the gray points indicating the initial planetesimals in our fiducial simulation.
	\label{fig:Mcrit_Disk1}
	}
\end{figure}

Figure \ref{fig:growth_Disk1} shows that it is not simple to define a given mass for which there will be no significant growth; given enough time and pebbles, even small planetesimals will eventually grow.  
Under the LJ2012 scenario, $\sim 4$ objects must grow to around $\sim 10 M_\oplus$, which means that at this location $M_{\rm drift} > 40 M_\oplus$.
Additionally, no other objects can grow significantly in the presence of this many pebbles, and thus they all must have initially been smaller than a few times $10^{-5} M_\oplus$. 

If we are interested in an equation to estimate this transition between growth and no growth, we can leverage the power of the exponential cutoff in Equation (\ref{eq:Rcap}).  
This suggests a critical cutoff of the form 
\begin{equation}
	M_{\rm crit} = \frac{1}{C^{1/\gamma}} \frac{t_s (\Theta\eta a\Omega)^3}{4 G}.
	\label{eq:Mcrit}
\end{equation}
If we choose $C=1$, this corresponds to the mass at which the mass accretion rate in Equation (\ref{eq:Rcap}) is reduced by a factor of $1/e$ (shown as a dotted line in Figure~\ref{fig:growth_Disk1}).
The dashed line shows the mass at which the mass accretion rate is reduced by a factor of $1/e^2$ ($C=2$).  
Given that we require little growth for objects below the cutoff when $M_{\rm drift} = 40 M_\oplus$, $C=2$ seems to be a reasonable choice as a critical curve.

In Figure \ref{fig:Mcrit_Disk1} we show the critical curves for various pebble sizes in the fiducial disk model.
For each pebble size we show our preferred critical curve (Equation~(\ref{eq:Mcrit}) with $C=2$). 
For comparison, these curves are overplotted on the initial planetesimals in our fiducial simulation.
There are a number of notable aspects of these curves.
First, the critical curve depends strongly on the size of the particle (or more explicitly, $t_s$).
Particles with large or small stopping times have critical growth masses orders of magnitudes smaller than particles with $\tau \sim 1$.
This means that if we have a range of pebble sizes then it is actually the extreme pebble sizes that contribute the most to the growth of the smaller planetesimals.  It is only once these planetesimals have grown that all of the pebbles are accreted efficiently.
Another interesting aspect to note about these critical curves is that, unless there is a transition from the Stokes to Epstein regime (which is the case for $s=0.5$m, 1 m), the critical mass increases with increasing distance from the star. We will return to this second observation in a moment.

\begin{figure} 
	\includegraphics[width=\columnwidth]{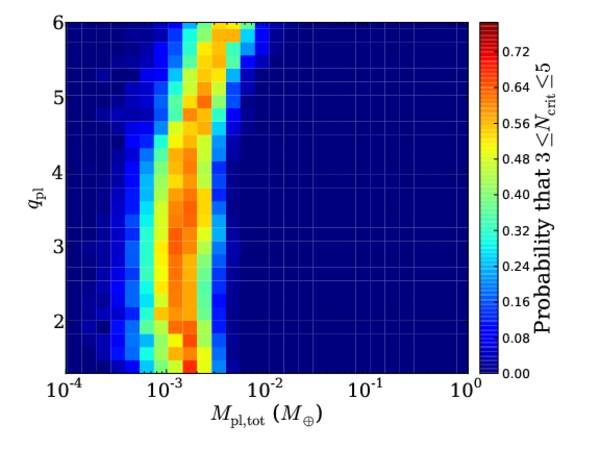}
	\caption{Probability that a randomly drawn planetesimal population with mass $M_{\rm pl,tot}$ and planetesimal slope $q_{\rm pl}$ will have three to five planetesimals above the critical curve.
	\label{fig:prob_hit}
	}
\end{figure}

With these curves in hand, we can evaluate the likelihood of LJ2012's idea that only four objects will grow.  In particular, we constructed Monte Carlo simulations where we placed a population of planetesimals in our fiducial $\Sigma \propto r^{-1}$ disk and used Equation (11) to determine which are larger than $M_{\rm crit}$.
We first randomly draw planetesimals from various size distributions where we vary the total mass ($M_{\rm pl, tot}$) and the slope ($q_{\rm pl}$).  Then we calculate the number of planetesimals greater than one of these critical curves.
Since we have a range of pebble sizes and the critical curves are different for each pebble mass, we define the relevant curve as being the lowest critical mass for all the pebbles in the distribution. 
We then calculate the probability that a given planetesimal population will have precisely three to five planetesimals greater than this critical curve.
In Figure \ref{fig:prob_hit} we show the percentage of randomly drawn populations that match the above criteria, if we assume that planetesimals are only formed within a range of 100--1000 km in size.
There is a relatively small region of parameter space for which there is a probability of successfully producing a handful of planetesimals that can grow.  While it is possible that our solar system was simply a lucky one that happened to be in this range, we find that constraint a little uncomfortable.

Perhaps more importantly, for all reasonable values of $q_{\rm pl}$ the planetesimal disk must be extremely low mass ($\sim 10^{-3} M_\oplus$) in order to avoid creating several tens (or more) of large planets.  As a comparison, the Kuiper Belt today is estimated to have a mass of 0.04--0.1 $M_\oplus$ \citep{Gladman.etal.2001}, and the Oort Cloud is expected to have a mass of at least an Earth mass \citep{Dones.etal.2004}.
This makes the low mass seem to be a challenging requirement.  
Therefore, without specific arguments for how a disk would self-regulate to produce the proper mass of planetesimals such that there are only a few planetesimals larger than the critical mass (which depends on both the disk structure and the amount of pebbles drifting through the disk) while still being consistent with the observed small-body reservoirs, we conclude that the idea that only four planetesimals could grow to become giant planets is highly unlikely, although of course we cannot rule it out.

\section{Additional Model Variations}\label{sec:other}
In the literature there have been several suggestions to enhance giant planet formation by pebble accretion. In the next subsections we address a few of these ideas.  We note that we have certainly not exhausted all possibilities related to these ideas, but as the reader will see, none of them have yet obviously presented a solution to the difficulties presented in this paper.

\subsection{Shielding} \label{sec:shield}
As noted by \citet{Morbidelli.Nesvorny.2012}, when planets get large, they can accrete a substantial fraction of the pebbles that cross their orbits.  
As a result, they suggested that such a planet might be able to effectively starve planets interior to them.
If the larger planetesimals could prevent the growth of their interior neighbors, then this would break down the assumption used in our analytic derivation that all planetesimals have access to the same pebble swarm and potentially 
lead to a system where only a small number of embryos grow.

Of course, this shielding is natively included in the \LIPAD\ simulations.  
However, because we begin the simulations with a substantial amount of pebbles intermixed with the planetesimals, it is difficult to ascertain to what extent shielding is occurring.  Therefore, in order to specifically investigate shielding, we construct a new set of runs without initial intermixed pebbles and only allow pebbles to drift in from beyond 10 AU.

\begin{figure} 
	\includegraphics[width=\columnwidth]{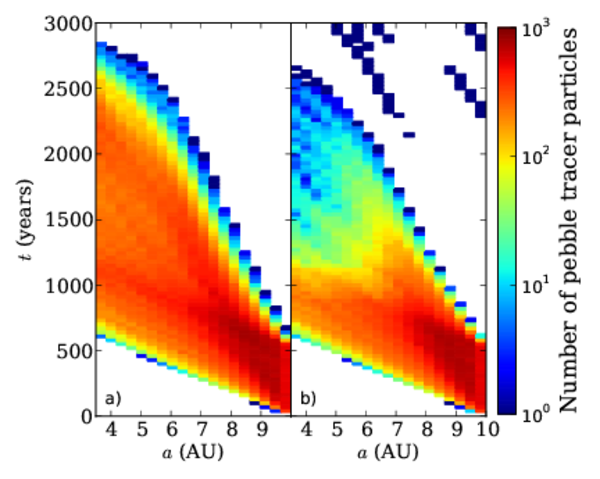}
	\caption{Number of pebble tracer particles as a function of location and time in our disk with only pebbles drifting in from the outside (similar to Figure \ref{fig:surfpebble_fid}).  In Panel (a) we show a simulation with no planetesimals; in Panel (b) we show our ``shielding'' run in which planetesimals are accreting the pebbles.  
	\label{fig:surfpebble_fid_shield}
	}
\end{figure}

\begin{figure}
	\includegraphics[width=\columnwidth]{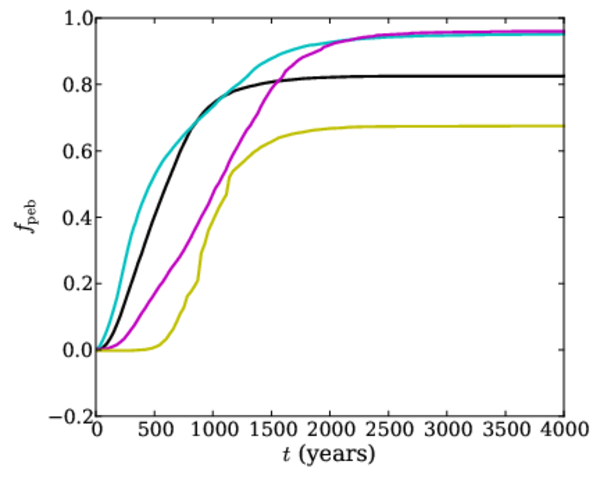}
	\caption{Total mass of planetesimals/planets in each simulation normalized by the total amount of solids that ever enter the computational domain.  The black curve is for the fiducial model and the cyan curve is for a similar simulation with large pebbles.  The yellow and magenta curves are for the shielding runs (in which there are no initial pebbles) with the fiducial pebbles and the large pebbles, respectively. 
	\label{fig:norm_eff_large}}
\end{figure}

Our first shielding run is identical to the fiducial model except for the lack of initial pebbles.
In Figure \ref{fig:surfpebble_fid_shield}(b) we show the distribution of pebbles as a function of time in our new shielding run.
This figure shows how pebbles drift inward from beyond 10 AU.
For ease of comparison, we also show the pebble surface density for the same pebble production, but without any planetesimals in Figure \ref{fig:surfpebble_fid_shield}(a).  It is clear that 
some pebbles are accreted as they drift inward, but a substantial fraction of the pebbles still reach the inner regions of the disk.
In the yellow curve in Figure \ref{fig:norm_eff_large} we show the fraction of the pebbles that have been accreted by the planetesimals as a function of time.
By the end, only 67\% of the pebbles have been accreted.
A comparison with our fiducial run (black curve) shows that \emph{more} pebbles are accreted there than in the shielding run.  This is because the planetesimals in the fiducial run have access to more pebbles in total so they grow larger, and as they grow larger they become more efficient at accreting pebbles.
But by the time the capture efficiency has gotten large, there already is a population of massive oligarchs, so shielding is not effective at alleviating the oligarchic problem in either the fiducial run or our shielding run.

However, as shown in Figure \ref{fig:Peff} and discussed by \citet{Morbidelli.Nesvorny.2012}, the efficiencies of pebble accretion depend strongly on the pebble properties.  By focusing on the 2D accretion regime, \citeauthor{Morbidelli.Nesvorny.2012} found that the efficiency of pebble accretion increases for particles with slow radial drift speeds; particles with either relatively large ($\tau \gg 1$) or small ($\tau \ll 1$) Stoke's numbers. 
If one allows that particles with $\tau \ll 1$ may be in the 3D accretion regime because they are easily excited to high scale heights in a moderately turbulent disk, then we would expect the capture efficiency to be the highest for large pebbles.  
This is related to Figure \ref{fig:Peff}; in our simple approximation the growth rate of large planets, when accreting large pebbles, is set by the rate at which those pebbles can be supplied to the growing planet.   While in a realistic simulation the capture efficiency is likely not actually unity, we would expect large planets to let very few (if any) large pebbles drift by without being accreted.

\begin{figure} 
	\includegraphics[width=\columnwidth]{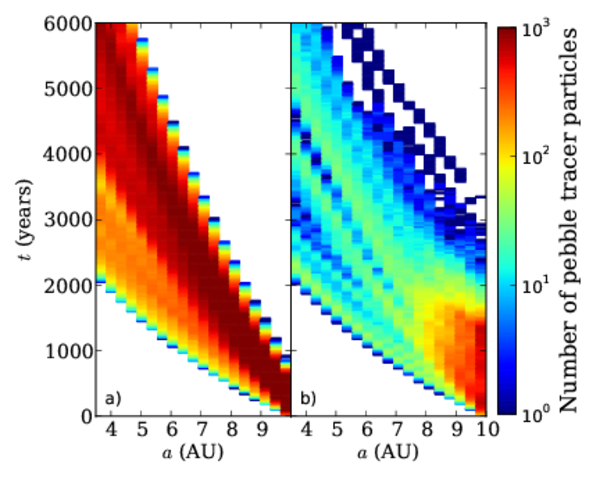}
	\caption{Number of pebble tracer particles as a function of location and time in our disk with only large pebbles drifting in from the outside.  See Figure~\ref{fig:surfpebble_fid_shield} for a description.
	\label{fig:surfpebble_shield}
	}
\end{figure}

To investigate shielding in a more favorable environment, we created runs in the fiducial disk with large ``pebbles'' (radii $s$=5--10 m).  As in the previous shielding experiment, we feed pebbles in only from the outside.
Figure \ref{fig:surfpebble_shield} shows the distribution of the pebbles in this simulation.  As compared to Figure \ref{fig:surfpebble_fid_shield}, it is clear that fewer pebbles are making it into the inner regions of the disk.  We can quantify this by comparing the number of pebbles that have crossed a given annulus with a test run in which we produce the same number of pebbles at 10 AU but have no planetesimals to accrete them.
Indeed, over 90\% of the pebbles are accreted when there are large pebbles.  As seen in Figure~\ref{fig:norm_eff_large}, the fraction is quite large both in a run in which large pebbles are intermixed (cyan curve) and in the shielding run in which large pebbles only drift in from beyond 10 AU (magenta curve).  For clarity we will discuss only this later run from this point forward.

\begin{figure} 
	\includegraphics[width=\columnwidth]{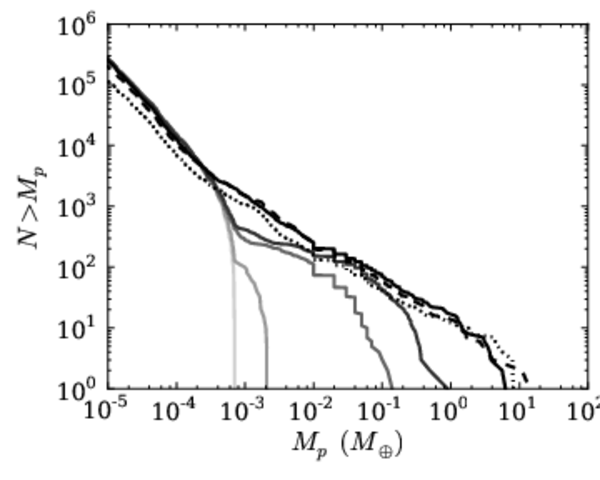}
	\caption{Evolving cumulative distribution when large pebbles are only coming from the outside.  The curves are the same as in Figure \ref{fig:fiducial}.
	\label{fig:sheilding}
	}
\end{figure}

Despite the high pebble accretion efficiencies in the large pebble shielding run, there is no sign of one (or a few) planetesimals running away to become giant planets in the cumulative size distribution (Figure~\ref{fig:sheilding}).
Indeed, the ineffectiveness of shielding is not that surprising if one looks at the strong mass dependence of the filtering fraction as seen in \citet{Morbidelli.Nesvorny.2012}.  
Initially, when the planetesimals are small, the filtering fraction of an individual planetesimal is not very large.  It is only when the bodies get relatively large (greater than 1 $M_\oplus$) that they sweep up a substantial fraction of the pebbles that cross their orbits.
In this simulation, no one planet becomes large enough to really ``starve'' their neighbors before they had a chance to grow.

\begin{figure} 
	\includegraphics[width=\columnwidth]{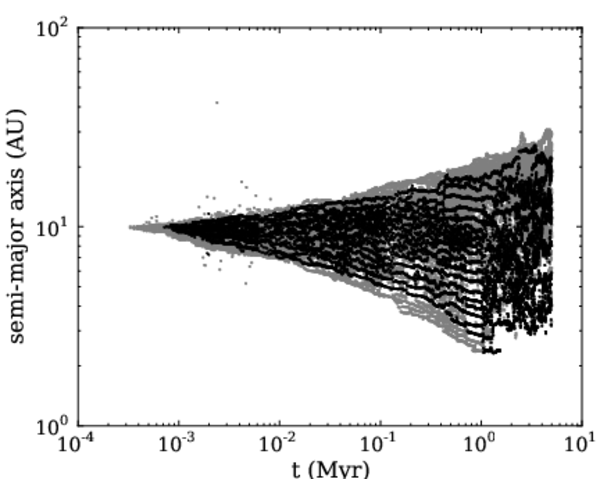}
	\caption{Semimajor axis as a function of time for large pebbles in the fiducial disk with pebbles only drifting in from the outside.  The symbols are the same as in Figure \ref{fig:fiducial_a}. 
	\label{fig:shielding_a}
	}
\end{figure}

However, from Figure~\ref{fig:surfpebble_fid_shield} it is clear that even if no individual planetesimal is capable of preventing the growth of neighbors, as an aggregate population the outer planetesimals do in fact accrete most of the pebbles, preventing the inner ones from growing.  
Thus, we might expect another possible mechanism by which shielding could aid in the growth of a single core; if all of the growing proto-cores are in the outer annulus of the disk, perhaps they could collide and merge into a single object.
However, as the cumulative size distribution clearly shows, this did not happen in our calculation.
By looking at the semimajor axis evolution in Figure \ref{fig:shielding_a}, we can see that though all of the planets did indeed form in the outer region of the disk, these planets scattered each other and spread.  This indicates that a high aggregate shielding rate is not sufficient to form a single core; for shielding to be effective, it must be for an individual object.  Even under these most favorable conditions the filtering fraction of individual planetesimals is not high enough to stop hundreds of Mars-mass and tens of Earth-mass planets from growing, and spreading, in this disk.
It appears that despite the higher accretion efficiency, shielding cannot provide a solution to this oligarchic problem.


\subsection{Presence of a Snow Line} \label{sec:snowline}
Another possible way to enhance growth in a localized region is by use of a snow line.  As icy pebbles drift inward, they will reach a point in the disk where they will begin to sublimate.  The volatiles will then diffuse both inward and outward, but inward-diffusing vapor can only be lost on the gas diffusion timescale, a timescale long compared to the pebble drift timescale while outward-diffusing vapor can be deposited on grains and return to the snow line on the short drift timescale.  This can significantly enhance the solid surface density on the outside of the snow line \citep{Stevenson.Lunine.1988, Cuzzi.Zahnle.2004}.
Recently, \citet{Ros.Johansen.2013} demonstrated that this effect is particularly effective at enhancing the growth of pebble-sized objects.
They demonstrated that condensation fronts can produce a significant amount of pebbles in a local region. 
They suggested that this mechanism could potentially produce an excess of pebbles, sufficient to allow pebble accretion to dominate in a narrow annulus.

\begin{figure} 
	\includegraphics[width=\columnwidth]{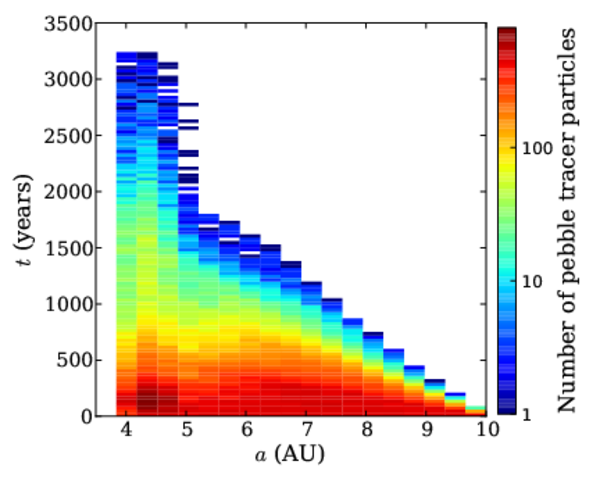}
	\caption{Distribution of pebbles as a function of time in a disk with an evaporation front.
	\label{fig:snowline_peb}
	}
\end{figure}

To investigate the above idea, we have run a variant of our fiducial model in which we place an evaporation front at 4 AU.  For simplicity, instead of reproducing the detailed pebble growth rates of \citeauthor{Ros.Johansen.2013}, we include a simple model to roughly mimic their conclusions.
We assume that 99\% of the pebbles that cross the evaporation front are recycled and reformed at a random distance between 4 and 5 AU.  The other 1\% of the mass is assumed to have remained in the gas phase and diffused into the inner solar system.
In this simulation we do not have pebbles coming in from beyond the computational domain, but we do start with pebbles intermixed between our planetesimals.
In Figure \ref{fig:snowline_peb} we show the number of pebble tracer particles as a function of location and time in the disk.  
One can see the enhancement of pebbles in the 4--5 AU region. 
Despite this high retention efficiency, all of the pebbles end up leaving the domain in just over 3000 yr, so overall the pebbles are not around significantly longer than in our standard simulations in which we add pebbles from the distant disk.

\begin{figure} 
	\includegraphics[width=\columnwidth]{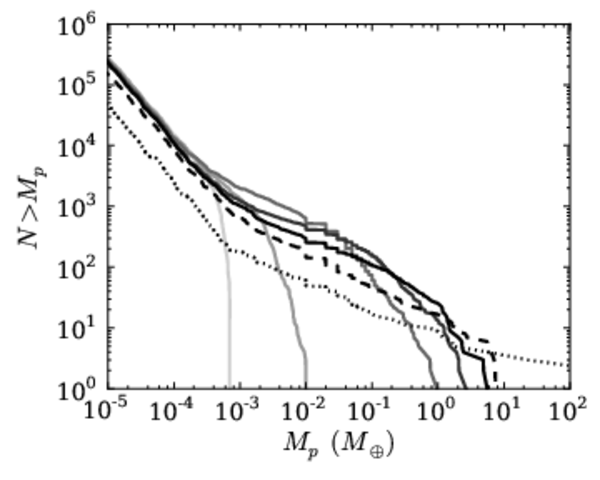}
	\caption{Evolving cumulative distribution in a disk with an evaporation front at 4 AU (as described in Section \ref{sec:snowline}).  The curves are the same as for Figure \ref{fig:fiducial}.
	\label{fig:snowline}
	}
\end{figure}

\begin{figure} 
	\includegraphics[width=\columnwidth]{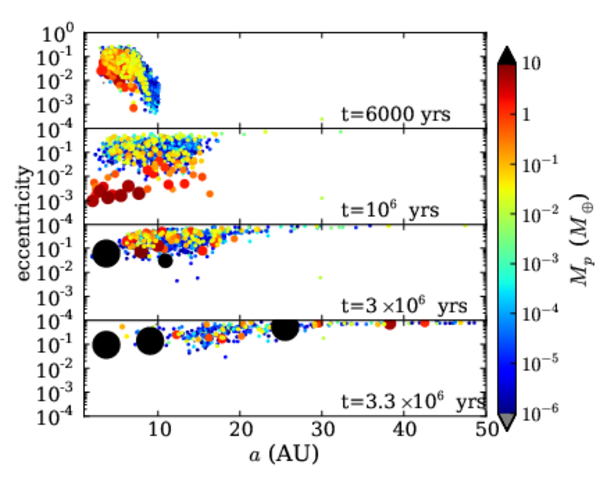}
	\caption{Snapshots from the model with an evaporation front at 4 AU. The symbols are the same as in Figure \ref{fig:fiducial_ae}.
	\label{fig:snowline_ae}
	}
\end{figure}

In Figures \ref{fig:snowline} and \ref{fig:snowline_ae} we see the evolution of the disk with the evaporation front in the same form as in prior figures.
The pebbles initially in the outer region drain inward very quickly, causing most of the growth to occur on the inner edge of the domain.
However, there is not just one core growing in the inner region; instead, a number of large cores compete to accrete the pebbles around the snow line.
These cores mutually excite each other, gaining high eccentricities.  This is related to the behavior that we saw in the shielding run; by forcing all of the pebble accretion to occur in a relatively narrow annulus, we do not prevent multiple planets from forming, but we do increase the violence of the dynamical instabilities once the planets have time to interact with one another.
In this simulation eventually three giant planets do grow, but the eccentricities of all these planets are above 0.1.
This again is not consistent with our own solar system.

\subsection{Presence of a Planet Trap} \label{sec:trap}
In the previous calculations we have neglected type I migration, the migration of embedded planets in a gaseous disk due to tidal interactions with the disk \citep{Ward.1997}.  
This migration poses a significant challenge to giant planet formation via core accretion.  Using the ``classical'' rate (such as that given in \citet{Tanaka.etal.2002}), the migration timescale for a 1 $M_\oplus$ core at 5 AU is less than $10^{6}$ yr in a MMSN, and decreases with increasing planet and disk mass.
This is shorter than the disk lifetime and thus has proved a substantial challenge for core formation.
One promising way to avoid losing the cores of the giant planets is to modify the disk structure to slow or even halt type I migration; such a location has been known in the literature as a planet trap \citep{Masset.etal.2006a}.
There are a number of theoretical reasons why planet traps might exist, particularly around the snow line.
For example, they could arise as a result of changes in the disk opacity 
\citep{Baruteau.Masset.2008, Paardekooper.Papaloizou.2008,Bitsch.Kley.2011,Bitsch.etal.2013} or changes in the disk viscosity \citep[][Bitsch et al. 2014, submitted]{Kretke.Lin.2007,Kretke.Lin.2012}.

We might expect that including type I migration and a planet trap would aid the core formation process by forcing our proto-cores close together, thus allowing them to merge and grow at a specific location.
Therefore, we performed simulations where, in addition to our snow line, we have added type I migration and a planet trap slightly interior to the snow line.  
We use the migration rate from \citet{Papaloizou.Larwood.2000} as described \citet{Levison.etal.2012}, setting the scaling factor in that paper's Equation (28), $c_a=1$.
We place a planet trap at 3.5 AU, interior to the snow line at 4 AU.  We have a very simple model of the trap: exterior to the trap type one migration is directed inward as normal, but for a few disk scale heights (0.2 AU) interior to the planet trap we multiply the fiducial migration rate by $-1$, so that the planet moves outward.
Within 3.3 AU we turn off migration.
This is not intended to be a high-fidelity reproduction of any specific planet trap model, 
but instead is a simple, computationally convenient model for preliminary studies.
Additionally, this represents an optimistic case because both planets of all masses are trapped and planets are not forced to continue migration inward if they are interior to the ``trap.''

\begin{figure} 
	\includegraphics[width=\columnwidth]{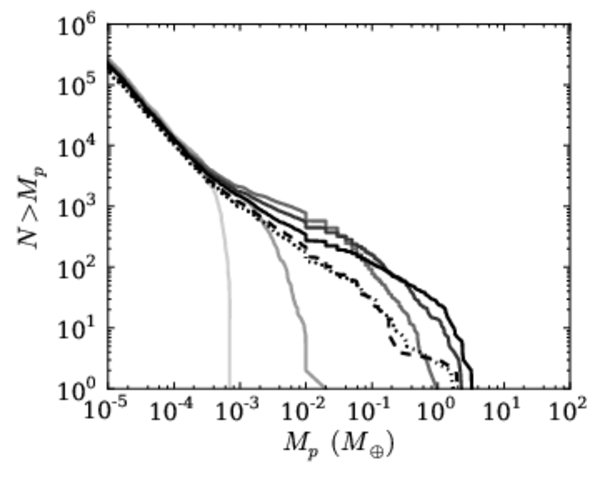}
\caption{Evolving cumulative distribution of a disk with a planet trap.  The curves are the same as in Figure~\ref{fig:fiducial}.
	\label{fig:planet_trap2}
	}
\end{figure}

\begin{figure} 
	\includegraphics[width=\columnwidth]{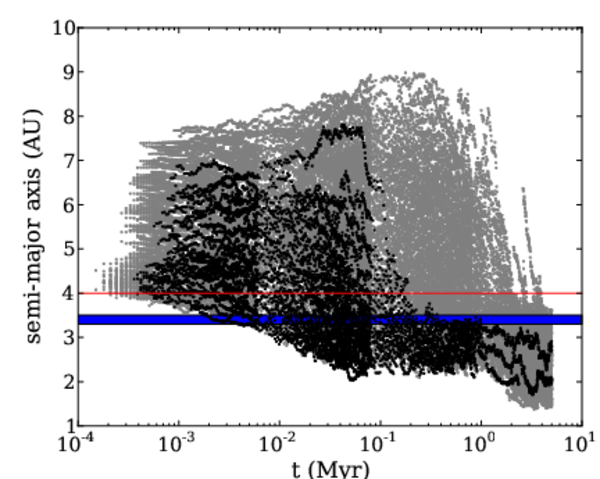}
	\caption{Semimajor axis of particles as a function of time in a disk with a planet trap.  The points are the same as in Figure \ref{fig:fiducial_a}.  The horizontal red line indicates the position of the snow line, while the blue shaded region indicates the region with outward migration.  Note that no protoplanets larger than $10 M_\oplus$ form in this simulation.
	\label{fig:planet_trap2_a}
	}
\end{figure}

Nevertheless, when we look at the cumulative distribution of the planet trap model (Figure \ref{fig:planet_trap2}), we see that planets never grow beyond a few Earth masses and there is significant loss of material within 5 Myr.
After 6000 yr tens of objects greater than 1 $M_\oplus$ have formed, but over the next couple million years of evolution all but two have been lost.
The reason for this loss becomes evident when one looks at the semimajor-axis evolution of the particles (Figure \ref{fig:planet_trap2_a}).  
Pebbles cause the planetesimals to grow to several Earth masses in the region between the snow line and 7 AU, and these all migrate inward as one would expect.  
However, once at the planet trap these embryos do not in general merge to form a single planet; instead, they scatter each other.  Planets that are scattered outward migrate back toward the trap; however, inwardly scattered planets can be delivered beyond the width of the trap and into the terrestrial planet region where there is no migration.
In this inner region planets continue to scatter off of each other (and the planets in the trap) allowing some planets to continue to spread inward.  All the while, planets that are scattered outward always return to the trap.
As a result, more and more icy planets are delivered to the realm of the terrestrial planets, which is clearly inconsistent with our solar system.
As we have placed the inner edge of the computational domain at 1 AU, we lose around 70 $M_\oplus$ of planets/planetesimals.  Therefore, while we cannot be sure of the final fate of these objects, it is clear that the presence of a planet trap does not force all of the embryos to merge into a single object in this model.

One may speculate that, if we had allowed type~I migration to continue in the
inner region, many of these ice-rich planets may have migrated on to the star,
leaving a dry terrestrial planet region.  While this is possible, it would
require two conditions to be met. First, the terrestrial planet building blocks
need to not migrate; therefore, they have to be small enough so that type~I
migration does not remove them and large enough not to drift inward from
aerodynamic drag.  Second, after the icy planets drifted through exciting their
eccentricities, they would need to damp in order to form planets.  Therefore,
they have to be either large enough to experience type I eccentricity damping
or small enough to experience significant gas drag.  It is not obvious that
there must exist a range of protoplanet sizes that can satisfy both of these
criteria and, if there is, that it will likely be small.  Therefore, we believe that the
disk would have to be very fine-tuned for the terrestrial planets to survive this type of migration.

As a side note, the ``leakiness'' of planet traps appears likely to be a recurring problem for any model of assembling giant planet cores at planet traps as long as the ``building block'' components are large enough to scatter each other outside of the trapping region.  For traps with widths a couple times the disk scale height, this corresponds to objects with escape velocities $\gtrsim 0.1$ the local Keplerian velocity (which is only roughly a Pluto mass at 5 AU).

\section{Discussion and Conclusions} \label{sec:discussion}
Given a population of planetesimals and pebbles (objects with stopping times comparable to their orbital times), pebble accretion is potentially an extremely rapid mechanism to grow planets.  However, despite the rapid growth rates, we find that invoking pebble accretion as the primary mechanism to form the cores of the giants planets is challenging.

As a test of the viability of pebble accretion for giant planet core formation in our solar system we preform a series of numerical simulations.  We follow the dynamical evolution of an initial population of $20~M_\oplus$ of planetesimals and 80 $M_\oplus$ of pebbles spread between 4 and 10 AU as they are allowed to collide, grow, fragment, and accrete pebbles.
We find that pebble accretion efficiently converts planetesimals with initial radii between 300 and 1000 km into Mars-, Earth-, or even larger-mass cores with only a relatively modest amount of pebbles drifting by.
However, pebble accretion tends to convert too many planetesimals into these large embryos.  

The crux of the issue is that although small planetesimals grow in a runaway manner (in which the mass doubling timescale decreases with planet mass), once protoplanets have entered the regime in which they are most efficiently accreting pebbles, their growth rate looks much more like the oligarchic stage in the standard planet formation scenarios \citep{Kokubo.Ida.1998, Kokubo.Ida.2000}, in which all of the planets grow together.
This is due to the fact that their effective capture radius only grows with the Hill sphere, $M_p^{1/3}$.  Additionally, once this large cross section is combined with the fact that pebbles tend to settle toward the midplane on a much shorter timescale than they drift inward \citep{Garaud.etal.2004}, accretion in the Hill regime tends to be two-dimensional.  Even with the fact that the Keplerian shear brings more material into the capture area of larger planets, the total growth rate only scales as $M_p^{2/3}$. 
If the oligarchs reach several Earth masses, then it is likely that a few will collide and allow some to grow large enough to accrete gas, but in addition to these ``lucky'' cores there will remain a large population of leftover ``oligarchs'' that violently dynamically interact with each other, spreading throughout the solar system and thereby polluting the terrestrial planet region with too much water-rich material and destroying any possibility of a cold population of planetesimals such as we see in our Kuiper Belt.

Changes to the disk structure cannot easily overcome this problem.  As one alters the disk structure in order to push the transition to oligarchic growth to higher masses, one must by necessity decrease the capture efficiency, driving up the needed mass of pebbles to unreasonably high levels.  Even if the initial disk was massive enough to create a sufficient quality of pebbles, with this method we require an extraordinarily steep disk profile ($\eta>3\times 10^{-2}$, which requires $\Sigma \propto < r^{-15}$ for standard temperature profiles), much steeper than the MMSN or observed disks \citep{Andrews.etal.2010}.
Moving to smaller, more excited pebbles to keep the growth closer to the runaway regime suffers a similar dilemma in that the efficiency of growth goes down significantly, requiring a very massive pebble reservoir, and still it is not enough to allow a small number of cores to run away to $10 M_\oplus$.

Additionally, it has been suggested that perhaps only four objects in our solar system were large enough to efficiently accrete pebbles, meaning that the oligarchic-type growth would not be problematic.  We find that a population of planetesimals consistent with the observed small-body reservoirs is very unlikely to have produced only a handful of planetesimals capable of growth (Section~\ref{sec:planetesimals}).
Furthermore, changing the parameters to increase the efficiency of shielding (Section~\ref{sec:shield}), adding sublimation fronts (Section~\ref{sec:snowline}), and even adding planet traps (Section~\ref{sec:trap}) do not appear to substantially aid the growth of a single (or a few) planetesimals at the expense of their neighbors.
Situations in which growth is concentrated in a relatively small annulus still have oligarchic growth within that annulus, and the proto-cores scatter each other out of the preferred region.
This demonstrates that the scattering and spreading that we see in the fiducial simulation are quite robust processes and hard to escape.
Of course, the calculations in the paper have not exhausted parameter space, and particularly our snow-line model and planet trap model are fairly simplistic.  Still, these runs show that it is not trivial to claim that a preferred region of planet formation automatically allows a single (or a few) giant planet cores to form.  The fundamental oligarchic nature of pebble accretion tends to win out.

While pebble accretion appears unavoidable if one creates planetesimals via
streaming instability, we find it to be challenging to use the sweeping up of
leftover pebbles as the primary mechanism to form our own solar system giant
planets.  Thus the solution to giant planet core formation remains elusive.

\acknowledgements
We would like to thank C.~Ormel, A.~Johansen, M.~Lambrechts, K.~Ros, K.~Walsh, and D.~Nesvorny for many stimulating discussions.  We would also like to thank J.~Chambers for his careful reading of the manuscript and useful comments.
This work was supported by grants from NSF AAG and NASA Origins (PI:~Levison).

\bibliography{copy}



\end{document}